%% file: paper_v14.tex
\documentclass[aps,prd,preprint,groupedaddress,amsmath,amssymb]{revtex4}
\usepackage{epsfig}

\begin{document}

\include{def}

\preprint{KA-TP-06-2012}
\preprint{LPSC 11-034}
\preprint{MS-TP-12-02}
\title{Associated production of charged Higgs bosons and top quarks with POWHEG}
\author{Michael Klasen$^a$}
\email[]{michael.klasen@uni-muenster.de}
\author{Karol Kova\v{r}\'{\i}k$^b$}
\email[]{kovarik@particle.uni-karlsruhe.de}
\author{Paolo Nason$^c$}
\email[]{paolo.nason@mib.infn.it}
\author{Carole Weydert$^d$}
\email[]{weydert@cern.ch}
\affiliation{$^a$ Institut f\"ur Theoretische Physik, Westf\"alische
 Wilhelms-Universit\"at M\"unster, Wilhelm-Klemm-Stra\ss{}e 9, D-48149 M\"unster,
 Germany\\
 $^b$ Institut f\"ur Theoretische Physik, Karlsruhe Institute of Technology,
 Postfach 6980, D-76128 Karlsruhe, Germany\\
 $^c$ INFN, Sezione di Milano-Bicocca, Piazza della Scienza 3, I-20126 Milan,
 Italy \\
 $^d$ Laboratoire de Physique Subatomique et de Cosmologie,
 Universit\'e Joseph Fourier/CNRS-IN2P3/INPG,
 53 Avenue des Martyrs, F-38026 Grenoble, France}
\date{\today}
\begin{abstract}
The associated production of charged Higgs bosons and top quarks at hadron
colliders is an important discovery channel to establish the existence of a
non-minimal Higgs sector. Here, we present details of a next-to-leading order
(NLO) calculation of this process using the Catani-Seymour dipole formalism and
describe its implementation in POWHEG, which allows to match NLO calculations
to parton showers. Numerical predictions are presented using the PYTHIA parton
shower and are compared to those obtained previously at fixed order, to a
leading order calculation matched to the PYTHIA parton shower, and to a
different NLO calculation matched to the HERWIG parton shower with MC@NLO. We
also present numerical predictions and theoretical uncertainties for various
Two Higgs Doublet Models at the Tevatron and LHC.
\end{abstract}
\maketitle



\section{Introduction}
\label{sec:1}

One of the most important current goals in high-energy physics is the discovery of
the mechanism of electroweak symmetry breaking. While this can be achieved, as in
the Standard Model (SM), with a single Higgs doublet field, giving rise to only 
one physical neutral Higgs boson, more complex Higgs sectors are very well
possible and in some scenarios even necessary. E.g., in the Minimal Supersymmetric
SM, which represents one of the most promising theories to explain the large
hierarchy between the electroweak and gravitational scales, a second complex Higgs
doublet is required by supersymmetry
with the consequence that also charged Higgs bosons should exist.

At hadron colliders, the production mechanism of a charged Higgs boson
depends strongly on its mass. If it is sufficiently light, it will be
dominantly produced in decays of top quarks, which are themselves copiously
pair produced via the strong interaction. Experimental searches in this channel
have been performed at the Tevatron by both the CDF \cite{Aaltonen:2009ke} and
D0 \cite{:2009zh} collaborations and have led to limits on the top-quark
branching fraction and charged Higgs-boson mass as a function of $\tan\beta$,
the  ratio of the two Higgs vacuum expectation values (VEVs), for various
Two Higgs Doublet Models (2HDMs). However, if the charged Higgs boson is heavier
than the top quark, it is dominantly produced in association with top quarks with
a semi-weak production cross section. The D0 collaboration have searched for
charged Higgs bosons decaying into top and bottom quarks in the mass range from
180 to 300 GeV and found no candidates \cite{Abazov:2008rn}. At the LHC,
the ATLAS (and CMS) collaborations have already excluded top-quark branching ratios to
charged Higgs bosons with masses of 90 (80) to 160 (140) GeV and bottom quarks above
0.03$-$0.10 (0.25$-$0.28) using 1.03 fb$^{-1}$ (36 pb$^{-1}$) of data taken at
$\sqrt{S}=7$ TeV \cite{Pelikan:2012fw,cms:lighthiggs}. At 14 TeV and
with an
integrated luminosity of 30 fb$^{-1}$, the discovery reach may be extended to
masses of about 600 GeV using also the decay into tau leptons and neutrinos
\cite{Flechl:2009zza,Kinnunen:2008zz}. It may then also become possible to
determine the spin and couplings of the charged Higgs boson, thereby identifying
the type of the 2HDM realized in Nature. Searches for pair-produced charged Higgs
bosons decaying into tau leptons and neutrinos, second generation quarks and
$W$-bosons and light neutral Higgs bosons have been performed at LEP and have led
to mass limits of $m_H>76.7$ (78.6) GeV for all values of $\tan\beta$ in Type-I
\cite{Abdallah:2003wd} (Type-II \cite{:2001xy}) 2HDMs, where only one (both)
Higgs doublet(s) couple to the SM fermions. Indirect constraints
from flavor-changing neutral currents (FCNCs) such as $b\to s\gamma$ can be
considerably stronger, e.g.\ $m_H>295$ GeV for $\tan\beta\geq2$ in the absence
of other new physics sources \cite{Misiak:2006zs}.

In this paper, we concentrate on the associated production of top quarks and
charged Higgs bosons at hadron colliders, which is of particular phenomenological
importance for a wide range of masses and models. Conversely, $s$-channel single,
pair, and associated production of charged Higgs bosons with $W$-bosons are less
favorable in most models. Isolation of this signal within large SM backgrounds,
e.g.\ from top-quark pair and $W$-boson associated production, and an accurate
determination of the model parameters require precise predictions that go beyond
the next-to-leading order (NLO) accuracy in perturbative QCD obtained
previously \cite{Zhu:2001nt,Plehn:2002vy,Berger:2003sm}. We therefore present
here details of our re-calculation of this process at NLO using the Catani-Seymour
(CS) dipole subtraction formalism \cite{Catani:1996vz,Catani:2002hc}. The virtual
loop and unsubtracted real emission corrections are then matched with a parton
shower (PS) valid to all orders in the soft-collinear region using the POWHEG
method \cite{Nason:2004rx,Frixione:2007vw} in the POWHEG BOX framework
\cite{Alioli:2010xd}. A similar calculation
has been presented 
using the Frixione-Kunszt-Signer (FKS) subtraction
formalism \cite{Frixione:1995ms} and matching to the HERWIG PS with the MC@NLO
method \cite{Weydert:2009vr}. Other new physics processes recently implemented
in MC@NLO include, e.g., the hadroproduction of additional neutral gauge bosons
\cite{Fuks:2007gk}. Unlike MC@NLO, POWHEG produces events with positive
weight, which is important when the experimental analysis is performed via trained
multivariate techniques. 
POWHEG can be easily
interfaced to both HERWIG \cite{Corcella:2000bw} and PYTHIA
\cite{Sjostrand:2006za} and thus does not depend on the Monte Carlo (MC) program
used for subsequent showering.

The remainder of this paper is organized as follows: In Sec.\ \ref{sec:2},
we present details of our NLO calculation of top-quark and charged Higgs-boson
production. We emphasize the renormalization of wave functions, masses, and
couplings, in particular the one of the bottom Yukawa coupling, in the virtual
loop amplitudes as well as the isolation and cancellation of soft and collinear
divergences with the Catani-Seymour dipole formalism in the real emission
amplitudes. The implementation in POWHEG is described in Sec.\ \ref{sec:3}. As the
associated production of top quarks with charged Higgs bosons is very similar to
the one with $W$-bosons \cite{Re:2010bp}, we concentrate here on the differences
of the two channels. We also emphasize the non-trivial separation of the
associated production from top-quark pair production with subsequent top-quark
decay in scenarios, where the charged Higgs boson is lighter than the top quark,
using three methods: removing completely doubly-resonant diagrams, subtracting
them locally in phase space, or including everything, so that top-quark pair
production with the subsequent decay of an on-shell top quark into a charged Higgs
boson is effectively included at leading order (LO). In Sec.\ \ref{sec:4}, we
present a detailed numerical comparison of the new POWHEG implementation to the
pure NLO calculation without PS \cite{Plehn:2002vy}, to a tree-level calculation
matched to the PYTHIA parton shower \cite{Alwall:2005gs}, and to the MC@NLO
implementation with the HERWIG PS \cite{Weydert:2009vr}. We also give numerical
predictions and theoretical uncertainties for various 2HDMs at the Tevatron and
LHC. Our conclusions are summarized in Sec.\ \ref{sec:5}.

\section{NLO calculation}
\label{sec:2}

\subsection{Organization of the calculation}

\begin{figure}
 \centering
 \epsfig{file=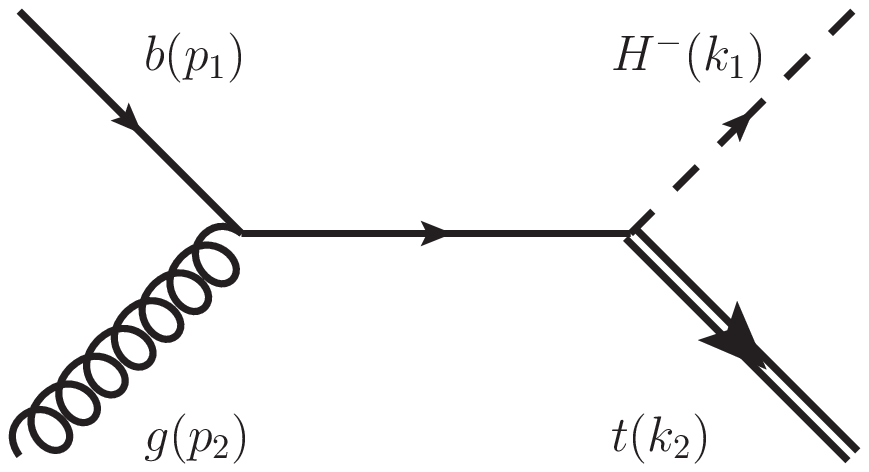,height=30mm}\qquad
 \epsfig{file=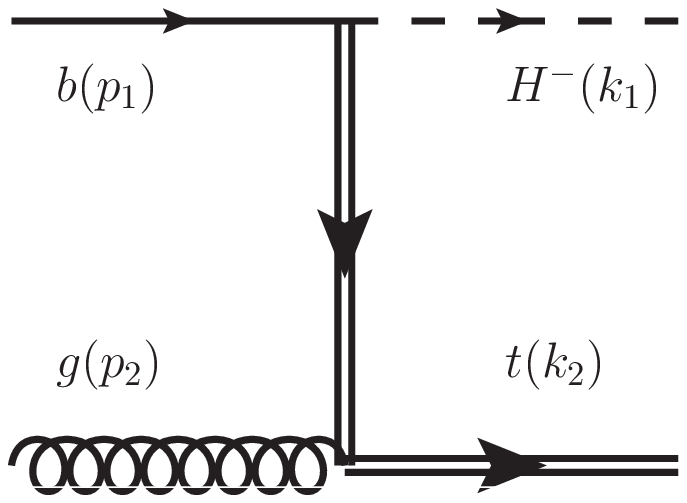,height=30mm}
 \caption{\label{fig:0}Tree-level diagrams for the associated production of
 charged Higgs bosons and top quarks at hadron colliders in the s-channel $\mathcal{S}$ and the t-channel $\mathcal{T}$.}
\end{figure}

At the tree level and in the five-flavor scheme with active bottom ($b$) quarks
as well as gluons ($g$) in protons and antiprotons, the production of charged
Higgs bosons ($H^-$) in association with top quarks ($t$) occurs at hadron
colliders via the process $b(p_1)+g(p_2) \rightarrow H^{-}(k_1)+t(k_2)$ through
the $s$- and $t$-channel diagrams shown in Fig.\ \ref{fig:0}. The massive top
%
%
quark is represented by a double line, whereas the bottom quark is treated
as massless and represented by a single line. The Born matrix elements can then
be given in terms of the Mandelstam variables 
\begin{eqnarray}
 s & = & (p_1+p_2)^2 = (k_1+k_2)^2, \\
 t & = & (p_2-k_2)^2 = (k_1-p_1)^2,~~{\rm and} \\
 u & = & m_t^2+m^2_H -s -t.
\end{eqnarray}
A NLO calculation in a four-flavor
scheme, where the bottom quark is treated as massive and generated by the
splitting of an initial gluon, has been presented elsewhere
\cite{Dittmaier:2009np}, but the effect of the bottom mass through the parton
densities was subsequently found to be strongly suppressed compared to its
impact on the bottom Yukawa coupling \cite{Plehn:2010gp}. The four-momenta 
of the participating particles have been ordered in accordance with the POWHEG
scheme, where the initial-state particles with four-momenta $p_1$ and $p_2$
are followed by the four-momentum $k_1$ of the final-state massive colorless
particle and then the four-momentum $k_2$ of the outgoing massive colored
particle. The additional radiation of a massless particle (gluon or light quark),
that occurs at NLO in real emission diagrams, is assigned the last four-momentum
$k_3$.

The hadronic cross section
\beq
 \sigma_{AB} = \sum_{a,b}\int_0^1dx_a \,f_{a/A}(x_a,\mu_F^2) \int_0^1dx_b\,
 f_{b/B}(x_b,\mu_F^2) \,\sigma(p_1,p_2;\mu_F^2)
 \label{eq:4}
\eeq
is obtained as usual as a convolution of the parton density functions (PDFs)
$f_{a/A,\,b/B}(x_{a,b},\mu_F^2)$ with the partonic cross section
\beq
 \sigma(p_1,p_2;\mu_F^2) = \sigma^{LO}(p_1,p_2)+\sigma^{NLO}(p_1,p_2;\mu_F^2)
\eeq
with partonic center-of-mass energy $s=x_ax_bS$, $S$ being the hadronic
center-of-mass energy. Its LO contribution
\beq
 \sigma^{LO}(p_1,p_2) = {1\over2s} \int d\Phi^{(2)}
 \overline{|\mathcal{M}_{\rm Born}|^2}
\eeq
is obtained from the spin- and color-averaged squared Born matrix elements
$\overline{|\mathcal{M}_{\rm Born}|^2}$ through integration over the two-particle
phase space $d \Phi^{(2)}$ and flux normalization.

\subsection{Virtual corrections and renormalization scheme}

Like the LO partonic cross section $\sigma^{LO}$, its NLO correction
\bea
 \label{all}
 \sigma^{NLO}(p_1,p_2;\mu_F^2) &=&
 \sigma^{NLO\{2\}}(p_1,p_2) + \sigma^{NLO\{3\}}(p_1,p_2)\nonumber\\
 &+&\int_0^1 dx \Bigl[ \sigma^{NLO\{2\}}(x;xp_1,p_2;\mu_F^2) +
 \sigma^{NLO\{2\}}(x;p_1,x p_2;\mu_F^2)\Bigr]
\eea
has a two-body final-state contribution
\bea
 \sigma^{NLO\{2\}}(p_1,p_2)
 &=&
 \int_{2} \left[  d\sigma^{V}(p_1,p_2) + d\sigma^{LO}(p_1,p_2)\otimes
 \mathbf{I} \right]_{\epsilon=0} \nonumber\\
 &=& \int d \Phi^{(2)} \biggl[ 2 \,\overline{\rm Re}\le
 \mathcal{M}_{\rm 1-loop}\,\mathcal{M}_{\rm Born}^\dagger\re
 +\,_{2}\langle H,t;b,g \mid \mathbf{I}(\epsilon)\mid H,t;b,g \rangle_{2}
 \biggr]_{\epsilon=0},\quad
 \label{all2}
\eea
which consists of the virtual cross section $d\sigma^V$, {\em i.e.} the spin-
and color-averaged interference of the Born diagrams with their one-loop
corrections, and the Born cross section $d\sigma^{LO}$ convolved with a
subtraction term $\mathbf{I}$, which can be written as $_{2} \langle
H,t;b,g \mid\mathbf{I}(\epsilon)\mid H,t;b,g \rangle_{2}$ and which removes the
infrared singularities present in the virtual corrections.
The three-body final-state contribution $\sigma^{NLO\{3\}}$ and the finite
remainders $\sigma^{NLO\{2\}}(x,...)$ of the initial-state singular terms will
be described in the third part of this section.

The ultraviolet divergencies contained in the virtual cross section
$d\sigma^{V}$ have been made explicit using dimensional regularization with
$D=4-2 \epsilon$ dimensions and are canceled against counterterms originating
from multiplicative renormalization of the parameters in the Lagrangian. In
particular, the wave functions for the external gluons, bottom and top quarks
are renormalized in the $\overline{\rm MS}$ scheme with
\bea
 \delta Z_g&=&-{\alpha_s\over4\pi}\le 2N_C-\lr{11\over3}N_C-{2\over3}N_F\rr\re
 \Delta_{UV} ~~~{\rm and} \\
 \delta Z_{b,t}&=&-{\alpha_s\over4\pi} C_F \Delta_{UV},
\eea
where $\Delta_{UV}=1/\epsilon-\gamma_E+\ln 4 \pi$, $\gamma_E$ is the Euler
constant, $N_C=3$ and $N_F=6$ are the total numbers of colors and quark flavors,
respectively, and $C_F = (N_C^2-1)/(2 N_C)$. The counterterm for the strong
coupling constant $\alpha_S=g_S^2/(4\pi)$
\beq
 {\delta g_S\over g_S} =
 -\frac{\alpha_S(\mu_R^2)}{8 \pi} \Bigl[ \Delta_{UV} \Bigl(
 \frac{11}{3} N_C - \frac{2}{3} N_F \Bigr) - \frac{2}{3} \ln \frac{\mu^2_R}{m_t^2}
 \Bigr],
\eeq
is computed in the $\overline{\rm MS}$ scheme using massless quarks, but
decoupling explicitly the heavy top quark with mass $m_t$ from the running of
$\alpha_S$ \cite{Collins:1978wz}.
The top-quark mass entering in the kinematics and propagators is renormalized in
the on-shell scheme,
\beq
 \frac{\delta m_t^{\rm OS} }{m_t} = -\frac{\alpha_S(\mu^2_R)}{4 \pi} 3 C_F \Bigl(
 \Delta_{UV} +\frac{4}{3} +\ln \frac{\mu^2_R}{m_t^2} \Bigr).
\eeq
On the other hand, we perform the renormalization of both the bottom and top
Yukawa couplings in the
$\overline{\rm MS}$ scheme,
\beq
 \frac{\delta y_{b,t}}{y_{b,t}(\mu^2_R)} =
 - \frac{\alpha_S (\mu^2_R)}{ 4 \pi} 3 C_F
 \Delta_{UV}
 .
 \label{eq:6}
\eeq
This enables us to factorize the charged Higgs-boson coupling at LO and NLO,
making the QCD correction ($K$) factors independent of the 2HDM and value of
$\tan\beta$ under study. In particular, in Eq.\ (\ref{eq:6}) we do not subtract the
mass logarithm, but rather resum it using the running quark masses
\beq
 \bar{m}_Q(\mu_R) = \bar{m}_Q(M_Q) \frac{c\Bigl( \alpha_s(\mu_R)/\pi\Bigr)}{c
 \Bigl( \alpha_s(M_Q)/\pi\Bigr)}
\eeq
in the Yukawa couplings, where
\beq
 c(x) = \Bigl( \frac{23}{6}x\Bigr)^{12/23} (1+1.175x+1.501x^2)
\eeq
for $m_b<\mu_R<m_t$ and
\beq
 c(x) = \Bigl( \frac{7}{2}x\Bigr)^{4/7} (1+1.398x+1.793 x^2)
\eeq
for $\mu_R > m_{b,t}$ \cite{Gorishnii:1990zu}.
The starting values of the $\overline{\rm MS}$ masses are
obtained from the on-shell masses $M_Q$ through the relation
\beq
 \bar{m}_Q(M_Q) = \frac{M_Q}{1+\frac{4}{3} \frac{\alpha_S(M_Q)}{\pi} + K_Q \Bigl(
 \frac{\alpha_s(M_Q)}{\pi}\Bigr)^2}
\eeq
with $K_b \approx 12.4$ and $K_t \approx 10.9$ \cite{Gray:1990yh,Djouadi:1997yw}.
 
After the renormalization of the ultraviolet singularities has been performed as
described above, the virtual cross section contains only infrared poles. These
are removed with the second term in Eq.\ (\ref{all2}), i.e.\ by
convolving the Born cross section with the subtraction term
\cite{Catani:1996vz,Catani:2002hc}
\beq
 \label{Iterms}
 \mathbf{I}(\epsilon) =
 \mathbf{I}_{2}(\epsilon,\mu^2;\{k_2,m_t \}) + \mathbf{I}_b (\epsilon,\mu^2;
 \{k_2,m_t \},p_1) +  \mathbf{I}_g (\epsilon,\mu^2;\{k_2,m_t \},p_2) +
 \mathbf{I}_{bg} (\epsilon,\mu^2;p_1, p_2),
\eeq
where in our case $\mathbf{I}_{2}(\epsilon,\mu^2;\{k_2,m_t \})=0$, since there are
no QCD dipoles with a final state emitter and a final state spectator. The dipoles
depending on one initial-state parton ($a=b,g$) with four-momentum $p_i$ ($i=1,2$)
are
\bea
 \label{IaGen}
 \mathbf{I}_a (\epsilon,\mu^2;\{k_2,m_t \},p_i) &=& - \frac{\alpha_s}{2 \pi}
 \frac{(4 \pi)^{\epsilon}}{\Gamma(1-\epsilon)} \lg \frac{1}{\mathbf{T}_t^2}
 \mathbf{T}_t \cdot \mathbf{T}_a  \le  \mathbf{T}_t^2
 \lr \frac{\mu^2}{s_{ta}}\rr^{\epsilon} 
 \lr \mathcal{V}_t (s_{ta},m_t,0;\epsilon)-\frac{\pi^2}{3} \rr \rp\rp\nonumber\\
 & & \hspace*{47mm} \lp +\,\Gamma_t(\mu,m_t;\epsilon) + \gamma_t \ln \frac{\mu^2}
 {s_{ta}} +\gamma_t + K_t \re\nonumber\\
 &+& \frac{1}{\mathbf{T}_a^2} \mathbf{T}_a \cdot \mathbf{T}_t \le
 \mathbf{T}^2_a \lr \frac{\mu^2}{s_{at}}\rr^{\epsilon}
 \lr \mathcal{V}_a (s_{at},0,m_t;\epsilon,\kappa)-\frac{\pi^2}{3}
 \rr \rp +\frac{\gamma_a}{\epsilon}  \nonumber\\
 & & \hspace*{18mm}\lp \lp  +\, \gamma_a \ln \frac{\mu^2}{s_{at}} +
 \gamma_a + K_a \re \rg,
\eea
where $\mathbf{T}_{a,t}$ denotes the color matrix associated to the emission of a
gluon from the parton $a$ or the top quark $t$, the dimensional regularization
scale $\mu$ is identified with the renormalization scale $\mu_R$, and $s_{ta}=
s_{at}=2p_ik_2$. The kernels
\bea
 \mathcal{V}_t(s_{ta},m_t,0;\epsilon) & = & \mathcal{V}^{(S)}(s_{ta},m_t,0;
 \epsilon)  +   \mathcal{V}_t^{(NS)}(s_{ta},m_t,0) \\
 \mathcal{V}_b (s_{bt},0,m_t;\epsilon,2/3) &  =  & \mathcal{V}^{(S)}
 (s_{bt},0,m_t;\epsilon)  + \mathcal{V}_b^{(NS)} (s_{bt},0,m_t) \\
 \mathcal{V}_g(s_{gt},0,m_t;\epsilon,2/3) & = &  \mathcal{V}^{(S)}
 (s_{gt},0,m_t;\epsilon) + \mathcal{V}_g^{(NS)} (s_{gt},0,m_t;2/3)
\eea
consist of the singular terms
\bea
 \mathcal{V}^{(S)}(s_{ta}\!\!\!&,&\!\!\!m_t,0;\epsilon) ~=~
 \mathcal{V}^{(S)}(s_{at},0,m_t; \epsilon) \nonumber\\
 &=& \frac{1}{2 \epsilon^2} + \frac{1}{2 \epsilon} \ln \frac{m_t^2}{s_{ta}}
 -\frac{1}{4} \ln^2 \frac{m_t^2}{s_{ta}}-\frac{\pi^2}{12}-\frac{1}{2} \ln
 \frac{m_t^2}{s_{ta}} \ln \frac{s_{ta}}{Q^2_{ta}}-\frac{1}{2} \ln \frac{m_t^2}
 {Q^2_{ta}} \ln \frac{s_{ta}}{Q^2_{ta}}
\eea
with $Q^2_{ta} = Q^2_{at} = s_{ta} + m_t^2 + m_a^2$ and the non-singular terms
\bea
 \mathcal{V}_t^{(NS)}(s_{ta},m_t,0) & = & \frac{\gamma_t}{\colm^2_t} \ln
 \frac{s_{ta}}{Q^2_{ta}}+\frac{\pi^2}{6}-\text{Li}_2 \biggl( \frac{s_{ta}}
 {Q^2_{ta}}\biggr)-2 \ln \frac{s_{ta}}{Q^2_{ta}}-\frac{m_t^2}{s_{ta}} \ln
 \frac{m_t^2}{Q^2_{ta}}, \\
 \mathcal{V}_b^{(NS)}(s_{bt},0,m_t) & = & \frac{\gamma_b}{\colm^2_b} \biggl[ \ln
 \frac{s_{bt}}{Q^2_{bt}}-2 \ln \frac{Q_{bt}-m_t}{Q_{bt}}-2\frac{ m_t}{Q_{bt}+m_t}
 \biggr] +\frac{\pi^2}{6} -\text{Li}_2 \biggl( \frac{s_{bt}}{Q^2_{bt}}\biggr),
 ~~ \\
 \mathcal{V}_g^{(NS)}(s_{gt},0,m_t;2/3)& = & \frac{\gamma_g}{\colm^2_g} \biggl[
 \ln \frac{s_{gt}}{Q^2_{gt}}-2 \ln \frac{Q_{gt}-m_t}{Q_{gt}}-2 \frac{m_t}{Q_{gt}+
 m_t} \biggr] +\frac{\pi^2}{6}-\text{Li}_2 \biggl( \frac{s_{gt}}{Q^2_{gt}} \biggr)
 \nonumber\\
 &+& \frac{4}{3} \frac{T_R}{N_C}\biggl[  \ln \frac{Q_{gt}-m_t}{Q_{gt}} +
 \frac{m_t}{Q_{gt}+m_t}-\frac{4}{3}   \biggr].
\eea
The constant $\kappa$ in Eq.\ (\ref{IaGen}) is a free parameter, which distributes
non-singular contributions between the different terms in Eq.\ (\ref{all}). The
choice $\kappa=2/3$ considerably simplifies the gluon kernel. For massive quarks,
one has in addition
\beq
 \Gamma_t(\mu,m_t;\epsilon) = C_F \biggl(  \frac{1}{\epsilon}+\frac{1}{2}\ln
 \frac{m_t^2}{\mu^2}-2 \biggr),
\eeq
while
\bea
 \gamma_q ~=~ \frac{3}{2} C_F ~~&,&~~
 \gamma_g ~=~ \frac{11}{6} N_C - \frac{2}{3} T_R N_f
\eea
and
\bea
 K_q ~=~ \biggl(  \frac{7}{2}-\frac{\pi^2}{6}\biggr) C_F ~~&,&~~
 K_g ~=~ \biggl( \frac{67}{18} -\frac{\pi^2}{6}\biggr)N_C -\frac{10}{9}T_R N_f
\eea
with $T_R=1/2$ and $N_f=5$ the number of light quark flavors.
The last term in Eq.\ (\ref{Iterms})
\bea
 \mathbf{I}_{bg} (\epsilon,\mu^2;p_1, p_2) &=& - \frac{\alpha_s}{2 \pi}
 \frac{(4 \pi)^{\epsilon}}{\Gamma(1-\epsilon)} \lg \frac{1}{\mathbf{T}_g^2}
 \mathbf{T}_g \!\cdot\!\colm_b \left[\lr\frac{\mu^2}{s_{bg}}\rr^{\epsilon}
 \lr\frac{\colm_g^2}{\epsilon^2}+\frac{\gamma_g}{\epsilon}\rr
 -\colm_g^2 \,\frac{\pi^2}{3}+\gamma_g + K_g \right] \rp \nonumber\\
 && \hspace*{25mm} +\,(g \leftrightarrow b) \biggl\}
\eea
depends on both initial-state partons.
Since the process we are interested in involves only three colored particles at
the Born level, the color algebra can be performed in closed form. To be concrete,
we have
\bea
 \mathbf{T}_{b} \cdot \mathbf{T}_{t} \vert H,t;b,g\ra_2 &=&
 -\left(C_F-\frac{N_C}{2}\right) \vert H,t;b,g\ra_2 = \frac{1}{2 N_C} \vert
 H,t;b,g\ra_2,\label{color1}\\
 \mathbf{T}_{b,t} \cdot \mathbf{T}_g \vert H,t;b,g\ra_2 &=&
 -\frac{N_C}{2} \vert H,t;b,g\ra_2,\label{color2}\\
 \mathbf{T}_{b,t} ^{2} \vert H,t;b,g\ra_2 &=&
 \quad C_F \,\vert H,t;b,g\ra_2,~~{\rm and}\\
 \mathbf{T}_{g} ^{2} \vert H,t;b,g\ra_2 &=& \quad N_C \ \vert H,t;b,g\ra_2.
\eea

\subsection{Real corrections}

The second term in Eq.\ (\ref{all})
\beq
 \sigma^{NLO\{3\}}(p_1,p_2)= \int d \Phi^{(3)} \Bigl\{
 \overline{|\mathcal{M}_{3,ij}(k_1,k_2,k_3;p_1,p_2)|^2}
 - \sum_{\text{dipoles}} \mathcal{D}(k_1,k_2,k_3;p_1,p_2) \Bigr\}
 \label{eq:30}
\eeq
includes the spin- and color-averaged squared real emission matrix elements
$\overline{|\mathcal{M}_{3,ij}(k_1,k_2,k_3;p_1, p_2)|^2}$ with three-particle
final states and the corresponding unintegrated QCD dipoles $\mathcal{D}$, which
compensate the integrated dipoles $\mathbf{I}$ in the previous section. Both terms
are integrated numerically over the three-particle differential phase space $d
\Phi^{(3)}$.

The real emission processes can be grouped into the four classes
\begin{enumerate}
\renewcommand{\labelenumi}{(\alph{enumi})}
\item $b(p_1)+g(p_2) \rightarrow H^{-}(k_1)+t(k_2)+g(k_3)$,
\item $g(p_1)+g(p_2) \rightarrow H^{-}(k_1)+t(k_2)+\bar{b}(k_3)$,
\item $b(p_1)+q/\bar{q}(p_2)\rightarrow H^{-}(k_1)+t(k_2)+q/\bar{q}(k_3) $,
 and
\item $q(p_1)+\bar{q}(p_2) \rightarrow H^{-}(k_1)+t(k_2)+\bar{b}(k_3)$,
\end{enumerate}
where the second process (b) can be obtained from the first one (a) by crossing
the four-momenta $k_3$ and $-p_1$ and multiplying the squared matrix element by a
factor of $(-1)$ to take into account the crossing of a fermion line. The
processes in the two other classes (c) and (d) can interfere when $q=b$, but these
contributions are numerically negligible due to the comparatively small bottom
quark parton distribution function. Process (d) is furthermore convergent
for $q=u,d,s$ and $c$.

The sum over the dipoles in Eq.\ (\ref{eq:30}) includes initial-state emitters
$ab$ with both initial- and final-state spectators $c$ ($\mathcal{D}^{ab,c}$ and
$\mathcal{D}^{ab}_c$) and the final-state emitter $ab$ with initial-state
spectators $c$ ($\mathcal{D}_{ab}^c$). For the three divergent processes, we have
\bea
 && (a) ~:~ \sum_{\text{dipoles}} = \mathcal{D}^{bg,g} +\mathcal{D}^{gg,b} + 
 \mathcal{D}^{bg}_t + \mathcal{D}^{gg}_t + \mathcal{D}_{tg}^b +
 \mathcal{D}_{tg}^g,\\
 && (b) ~:~ \sum_{\text{dipoles}} = \mathcal{D}^{g_1b,g_2} + \mathcal{D}^{g_2b,
 g_1} + \mathcal{D}^{g_1b}_t  + \mathcal{D}^{g_2b}_t,~~{\rm and}\\
 && (c) ~:~ \sum_{\text{dipoles}} = \mathcal{D}^{qq,b} + \mathcal{D}^{qq}_t.
\eea
Denoting by $a$ the original parton before emission, $b$ the spectator, and $i$
the emitted particle, the dipole for initial-state emitters and initial-state
spectators is given by
\beq
 \mathcal{D}^{ai,b} = -\frac{1}{2 p_a k_i} \frac{1}{x_{i,ab}}
 \; _{2,ab} \langle \tilde{H},\tilde{t}; \tilde{ai}, b \mid
 \frac{\mathbf{T}_b \cdot \mathbf{T}_{ai}}{\mathbf{T}_{ai}^2}
 \mathbf{V}^{ai,b}\mid \tilde{H},\tilde{t};\tilde{ai}, b \rangle_{2,ab},
\eeq
where the momentum of the intermediate initial-state parton $\tilde{ai}$ is
$\tilde{p}^{\mu}_{ai}= x_{i,ab}\,p_a^{\mu}$ with $x_{i,ab}=(p_a p_b - k_i p_a -
k_i p_b)/(p_a p_b)$, the momentum $p_b$ is unchanged, and the final-state momenta
$k_j$ with $j=1,2$ are shifted to
\beq
 \tilde{k}_{j}^{\mu} = k_{j}^{\mu}-\frac{2 k_j \cdot (K+\tilde{K})}
 {(K+\tilde{K})^2}(K+\tilde{K})^{\mu}+ \frac{2 k_j \cdot K}{K^2} \tilde{K}^{\mu}
\eeq
with $K^{\mu}= p_a^{\mu}+p_b^{\mu}-k_i^{\mu}$ and $\tilde{K}^{\mu}=\tilde{p}_{ai}
^{\mu}+ p_b^{\mu}$. The necessary splitting functions $\mathbf{V}^{ai,b}$ for
$\{ai,b\}=\{qg,g;gg,q;gq,g;qq,q\}$ can be found in Ref.\
\cite{Catani:1996vz}.
The dipole for initial-state emitters and a final-state spectator, which is in our
case the top quark $t$, is given by
\beq
 \mathcal{D}_t^{ai} = - \frac{1}{2 p_a k_i}
 \frac{1}{x_{it,a}} \; _{2,\tilde{ai}} \langle H,\tilde{t};\tilde{ai},b
 \mid \frac{\mathbf{T}_t \cdot \mathbf{T}_{ai}}{\mathbf{T}^2_{ai}}
 \mathbf{V}^{ai}_t \mid H,\tilde{t};\tilde{ai},b\rangle_{2,\tilde{ai}},
\eeq
where the momentum of the intermediate initial-state parton $\tilde{ai}$ is
$\tilde{p}_{ai}^{\mu} = x_{it,a} p_a^{\mu}$ with $x_{it,a} = (p_a k_i +
p_a p_t - k_i p_t)/(p_a k_i + p_a p_t)$, the momentum $p_b$ is
unchanged, and the momentum of the final-state top quark $p_t$ is shifted to
$\tilde{p}_{t}^{\mu} = k_i^{\mu} + p_t^{\mu}-(1-x_{it,a})p_a^{\mu}$. The
necessary splitting functions $\mathbf{V}^{ai}_t$ for $\{ai,t\}=\{qg,t;gg,t;
gq,t;qq,t\}$ can be found in Ref.\ \cite{Catani:2002hc}.
%
%
%
%
Finally, the dipole for final-state emitter (the top quark $t$) and initial-state
spectator $a$ is given
by
\beq
 \mathcal{D}^a_{tg} = -\frac{1}{2 p_t k_i}
 \frac{1}{x_{it,a}}\; _{2,a} \langle H,\tilde{it}; \tilde{a},b \mid
 \frac{\mathbf{T}_a \cdot \mathbf{T}_{it}}{\textbf{T}_{it}^2} \mathbf{V}^a_{it}
 \mid H, \tilde{it};\tilde{a},b\rangle_{2,a},
\eeq
where the momentum of the initial parton $a$ is shifted to $\tilde{p}_{a}^{\mu} =
x_{it,a}p_a^{\mu}$ with $x_{it,a} = (p_a k_i + p_a p_t - k_i p_t)/(p_a k_i + p_a
p_t)$, the momentum $p_b$ is unchanged, and the momentum of the intermediate
final-state top quark $p_t$ is $\tilde{p}_{it}^{\mu} = k_i^{\mu} + p_t^{\mu}-
(1-x_{it,a}) p_a^{\mu}$. The required splitting function $\mathbf{V}^a_{gt}$ can
again be found in Ref.\ \cite{Catani:2002hc}.
%

The last terms in Eq.\ (\ref{all}) are finite remainders from the cancellation of
the $\epsilon$-poles of the initial-state collinear counterterms. Their general
expressions read
\bea
 \int_0^1\!\!\!\!&&\!\!\!\!dx \,\sigma^{NLO\{2\}}\left(x;x p_1, p_2;\mu_F^2\right)
 ~=~ \sum_{a^{ \prime}} \int_0^1 dx \int_2 \left[  d \sigma^{LO}_{a^{\prime} b}
 \left( x p_1, p_2\right) \otimes \left( \mathbf{K}+\mathbf{P}\right)^{a,
 a^{\prime}} \left(x\right)   \right]_{\epsilon=0} \\
 &=&  \sum_{a^{\prime}} \int_0^1 \!dx \int \!d\Phi^{(2)}(xp_1,p_2)\
  _{2,a^{\prime} b} \langle k_1,k_2;xp_1,p_2\vert \mathbf{K}^{a,a^{\prime}}
 (x)+ \mathbf{P}^{a,a^{\prime}}(x;\mu_F^2) \vert k_1,k_2;xp_1,p_2
 \rangle _{2,a^{\prime} b}\nonumber
\eea
and similarly for $(a\leftrightarrow b)$ and $(p_1 \leftrightarrow p_2)$. The
color-charge operators $\mathbf{K}$ and $\mathbf{P}$ are explicitly given in Ref.\
\cite{Catani:2002hc}.

\section{POWHEG implementation}
\label{sec:3}

The calculation in the previous section has been performed using the
Catani-Seymour dipole formalism for massive partons \cite{Catani:1996vz,%
Catani:2002hc}.
For the implementation of our NLO calculation in the POWHEG Monte Carlo program,
we need to retain only the Born process, the finite terms of the virtual
contributions, and the real emission parts of our calculation, since all necessary
soft and collinear counterterms and finite remnants are calculated automatically
by the POWHEG BOX in the FKS scheme \cite{Frixione:1995ms}. Soft and collinear
radiation is then added to all orders using the Sudakov form factor. In this
section, we briefly describe the three relevant contributions, following closely
the presentation in Ref.\ \cite{Alioli:2010xd}, and address the non-trivial
separation of the associated production of charged Higgs bosons and top quarks
from top-quark pair production with subsequent top-quark decay in scenarios, where
the charged Higgs boson is lighter than the top quark.

\subsection{Born process}

In the POWHEG formalism, a process is defined by its particle content. Each
particle is encoded via the Particle Data Group numbering scheme
\cite{Nakamura:2010zzi} except for
gluons, which are assigned the value zero. The order of the final state particles
has to be respected. Colorless particles are listed first, then heavy colored
particles, and finally massless colored particles. The Born processes (two with
respect to the different $bg$ and $gb$ initial states) are defined with
${\tt flst\_nborn}=2$ and are listed as
\begin{equation}
 (bg \rightarrow H^-t) = \left[5,0,-37,6\right]
\end{equation}
and
\begin{equation}
 (gb \rightarrow H^-t) = \left[0,5,-37,6\right]
\end{equation}
in the subroutine \texttt{init\_processes}.

In the subroutine \texttt{born\_phsp}, the integration variables {\tt xborn(i)}
for the Born phase space are generated between zero and one. The hadronic cross
section is then obtained from the differential partonic cross section $d\sigma$
via the integration (see Eq.\ (\ref{eq:4}))
\begin{eqnarray}
 \sigma_{AB} & = & \sum_{a,b}\int_{0}^{1} d x_a f_{a/A}  \int_{0}^{1} d x_b
 f_{b/B}   \int_{t_{min}}^{t_{max}} \frac{d \sigma}{dt} dt \nonumber\\
& = & \int_{\tau_{min}}^{\tau_{max}} d \tau \int_{y_{min}}^{y_{max}} d y \; f_{a/A} \; f_{b/B} \int_{t_{min}}^{t_{max}} \frac{d \sigma}{dt} dt,
\end{eqnarray}
where $f_{i/I}$ is the PDF of parton $i$ inside hadron $I$ with momentum fraction
$x_i$ and where we have performed the change of variables
\bea
 y~=~ \ln \frac{x_a}{\sqrt{x_a x_b}} &~~~{\rm and}~~~& \tau ~=~ x_a x_b. 
\eea
The integration limits are given in Tab.\ \ref{intbord}.
\begin{table}
\centering
 \caption{\small{Integration limits for the hadronic cross section.\small{\label{intbord}}}}
\begin{tabular}{|c|c|c|}
\hline
Variable $V$& $V_{min}$ & $V_{max}$ \\
\hline
$\tau$ & $ \frac{(m_H+m_t)^2}{S}$ & 1 \\
$y$ & $\frac{1}{2} \ln \tau$ & $-\frac{1}{2} \ln \tau$\\
$t$ & $\frac{1}{2}(t_1 - t_2)$ & $\frac{1}{2}(t_1 + t_2)$\\
\hline
\multicolumn{3}{|c|}{} \\
\multicolumn{3}{|c|}{$t_{1}= m_t^2+m_H^2-s, \;t_{2}= \sqrt{(s-m_t^2-m_H^2)^2-4 m_t^2 m_H^2}$}\\
\multicolumn{3}{|c|}{} \\
\hline 
 \end{tabular}
\end{table}
The Jacobian for the change of integration variables from {\tt xborn(i)} to
$(\tau,y,t)$
\begin{equation}
 \Delta_{jac}= (\tau_{max}-\tau_{min})\times(y_{max}-y_{min})\times(t_{max}-t_{min})
\end{equation}
has to be multiplied with $2 \pi$ for the integration over the azimuthal angle
$\phi$, which is randomly generated by POWHEG. The different kinematical
variables can then be constructed in the center-of-mass reference frame as well as
in the laboratory frame via boosts.
The renormalization scale $\mu_R$ and factorization scale $\mu_F$ are set in the
subroutine \texttt{set\_fac\_ren\_scales} according to the usual convention
\begin{equation}
 \mu_R=\mu_F= \frac{m_t+m_H}{k},
\end{equation}
where $k$ is to be varied around two for uncertainty studies. Both the
\texttt{born\_phsp} and the \texttt{set\_fac\_ren\_scales} subroutines can be
found in the file \texttt{Born\_phsp.f}.

\begin{figure}
 \centering
\includegraphics[scale=0.5]{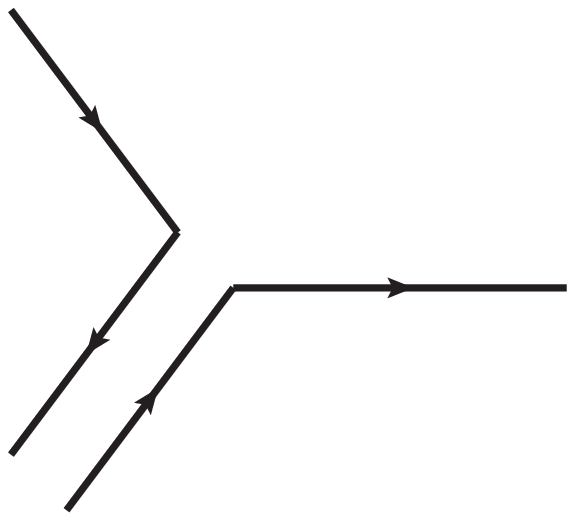} \hspace{3 cm}
\includegraphics[scale=0.5]{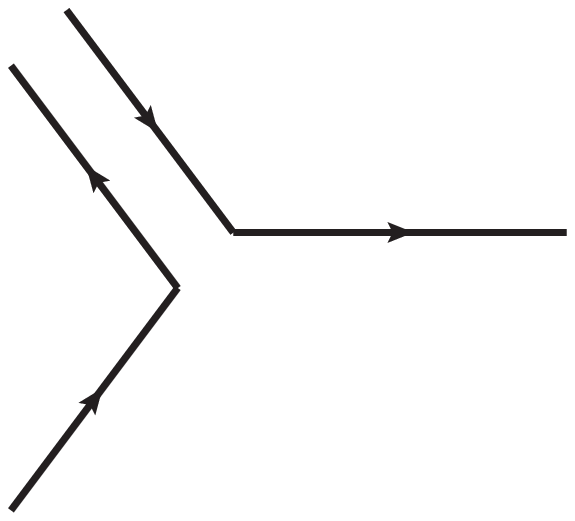}
\caption{\small{Color flow in the Born contribution $\left[5,0,-37,6\right]$ and for switched incoming partons $\left[0,5,-37,6\right].$\label{colconn}}}
\end{figure}

All other routines relevant to the Born process are contained in the file
\texttt{Born.f}. The subroutine \texttt{setborn} contains the factors for the
color-correlated Born amplitudes, which are related to the Born process through
the color factors quoted in Eqs.\ (\ref{color1}) and (\ref{color2}). The
subroutine \texttt{borncolor\_lh} contains the color flow of the Born term in the
large-$N_C$ limit shown in Fig.\ \ref{colconn}.
The routine \texttt{compborn} contains the spin-correlated Born matrix element
\begin{equation}
\mathcal{M}^{\mu\nu}_{\rm Born} = - \Bigl( \mathcal{S}^{\mu}\mathcal{S}^{\nu}+\mathcal{S}^{\mu} \mathcal{T}^{\nu}+ \mathcal{T}^{\mu}\mathcal{S}^{\nu}+ \mathcal{T}^{\mu}\mathcal{T}^{\nu} \Bigr)
\end{equation}
before summing over the initial gluon polarizations as well as
\begin{equation}
\mathcal{M}_{\rm Born}= - g_{\mu\nu}{\cal M}^{\mu\nu}_{\rm Born},
\end{equation}
where $g_{\mu\nu}$ is the metric tensor.

\subsection{Virtual loop corrections}

The renormalized virtual cross section is defined in dimensional regularization
and in the POWHEG convention by
 \begin{equation}\label{virt}
  \mathcal{V}= \frac{(4 \pi)^{\epsilon}}{\Gamma(1-\epsilon)}\Bigl( \frac{\mu_R^2}{Q^2}\Bigr)^{\epsilon} \frac{\alpha_s}{2 \pi} \Bigl[ \Bigl(\frac{C_2}{\epsilon^2}+\frac{C_1}{\epsilon} \Bigr)|\mathcal{M}_{\rm Born}|^2 + \mathcal{V}_{fin}\Bigr],
 \end{equation}
where $|\mathcal{M}_{\rm Born}|^2$ is now the squared Born matrix element computed
in $D=4-2\epsilon$ dimensions and where the remaining double and simple infrared
poles are proportional to
\begin{eqnarray}
 C_2 & = & \frac{1}{2 N_C}-\frac{3}{2} N_C ~~~{\rm and}\\
 C_1 &=& \frac{1}{4 N_C} \Bigl( 5-4 \ln \frac{m_t^2-u}{m_t^2}\Bigr)
 + \frac{N_C}{12} \Bigl(  -37 + 12 \ln \frac{s}{m_t^2} + 12 \ln \frac{m_t^2-t}{m_t^2}\Bigr) + \frac{1}{3} N_F.
\end{eqnarray}
The POWHEG implementation needs then only the finite coefficient
$\mathcal{V}_{fin}$, which has been organized into terms stemming from scalar 2-, 3- and
4-point integral functions $B_0$, $C_0$ and $D_0$ plus remaining terms and can be
found in the file \texttt{virtual.f}. Non-divergent $C_0$-functions and Euler
dilogarithms are computed using routines contained in the file \texttt{loopfun.f}.

\subsection{Real emission corrections}

In the subroutine \texttt{init\_processes}, the index of the first colored light
parton in the final state is defined, which is in our case the additional jet from
the real emission (${\tt flst\_lightpart}=5$). All ${\tt flst\_nreal}= 30$
real emission processes are then assigned a number according to the list given in
Tab.\ \ref{reallist}.
\begin{table}
\centering
\caption{\small{Process numbers of the different real emissions. Here $q=d,u,s,c.$ \label{reallist}}}
 \begin{tabular}{|cc|cc|}
\hline
Process number & Initial state & Process number & Initial state \\
\hline
1 & $bg$ & 16-19 & $\bar{q}b$ \\
2 & $gb$ & 20-23 & $q\bar{q}$ \\  
3 & $gg$ & 24-27 & $\bar{q}q$ \\
4-7 & $bq$ & 28 & $b \bar{b}$ \\
8-11 & $qb$ & 29 & $\bar{b}b$ \\
12-15 & $b \bar{q}$ & 30 & $bb$ \\
\hline  
 \end{tabular}
\end{table}
The expressions of the squared real emission matrix elements are given in the file
\texttt{real\_ampsq.f}.

\subsection{Separation of associated production and pair production of top quarks}

\begin{figure}
\centering
  \includegraphics[scale=0.4]{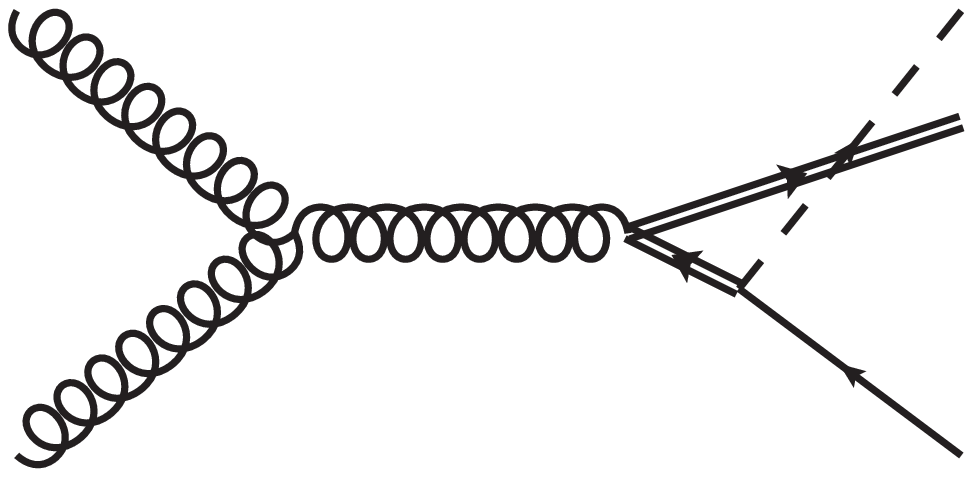} \hspace{0.2 cm} \includegraphics[scale=0.5]{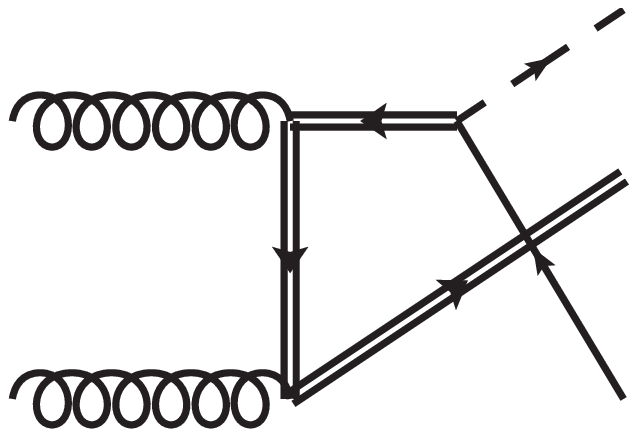} \hspace{0.2 cm}  \includegraphics[scale=0.55]{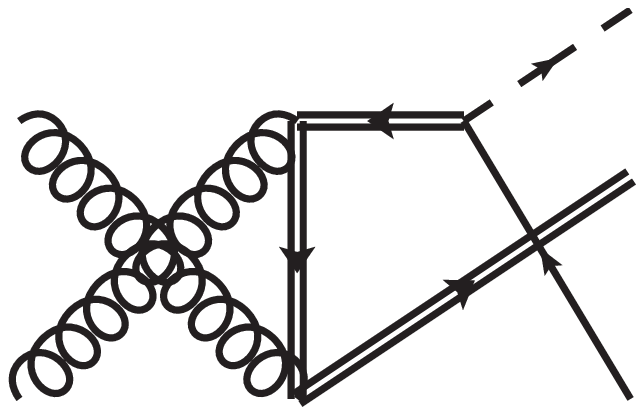}\hspace{0.2 cm}  \includegraphics[scale=0.5]{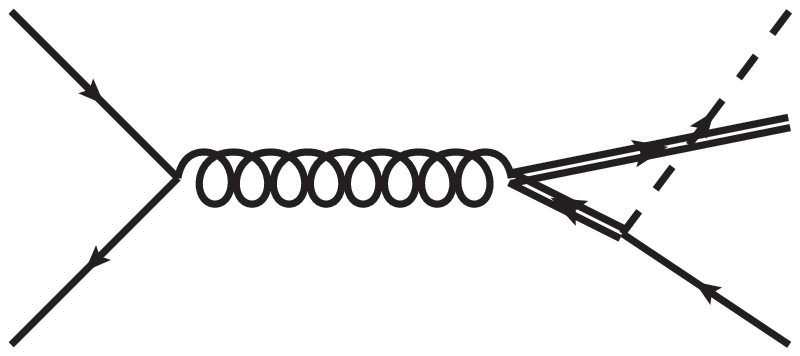}
\caption{\small{Real emission contributions in the gluon-gluon and quark-antiquark
 channels with an antitop propagator that can go on shell.\label{resdiag}}}
\end{figure}

If the charged Higgs-boson mass $m_H$ is lower than the top-quark mass
$m_t$, the antitop propagator of the real emission amplitudes shown in Fig.\
\ref{resdiag} can go on shell, resulting in a drastic increase of the total
cross section.  
In other words, the prevalent production mechanism becomes the on-shell
production of a $t\bar{t}$ pair, followed by the decay of the antitop quark
into a charged Higgs boson. The corresponding Feynman graphs contribute to
top-antitop production at LO with the charged Higgs boson being
produced in top-quark decays, but also to $tH^-$ production at NLO. The relevant
NLO processes are free from collinear and soft singularities.

At this point the problem arises how to separate the two production
mechanisms. In the literature two methods have been proposed: Diagram Removal (DR)
and Diagram Subtraction (DS) \cite{DRDS:2008}. Both remove the top-quark resonance
from the cross section, but the procedure for combining top pair production with
the associated production is not completely clear. If we separate the amplitudes
of a real emission process with colliding partons $a$ and $b$ into contributions
$\mathcal{M}_{ab}^{t \bar{t}}$, which proceed through $t\bar{t}$-production, and
contributions $\mathcal{M}_{ab}^{tH^-}$, which do not,
\begin{equation}
 \mathcal{M}_{ab}= \mathcal{M}_{ab}^{t \bar{t}} +
 \mathcal{M}_{ab}^{tH^-},
\end{equation}
squaring the amplitudes gives rise to three different quantities:
\begin{eqnarray}
 \vert \mathcal{M}_{ab} \vert^2  & = &  \vert \mathcal{M}_{ab}^{tH^-} \vert^2 + 2 {\rm Re} \bigl(\mathcal{M}_{ab}^{tH^-} \mathcal{M}_{ab}^{t \bar{t}*} \bigr) +  \vert \mathcal{M}_{ab}^{t \bar{t}} \vert^2
~ = ~ \mathcal{S}_{ab} + \mathcal{I}_{ab} + \mathcal{D}_{ab}.
\end{eqnarray}
The term $\mathcal{D}_{ab}$ contains neither collinear nor soft singularities,
while the interference term $\mathcal{I}_{ab}$ contains integrable infrared
singularities. These terms are therefore sometimes referred to as subleading with
respect to those in $\mathcal{S}_{ab}$, which contains all infrared singularities
and must be regularized, e.g., via the subtraction formalism.
DR requires removing $t\bar{t}$ production at the amplitude level. The only
contributing element is then $\mathcal{S}_{ab}$. Since it contains all
divergencies, the dipoles used in the $m_H>m_t$ case remain valid.
In the DS scheme, one subtracts from the cross section the quantity
\begin{equation}
d \sigma^{\rm sub}_{H^-t}=\frac{f_{\rm BW}(m_{H^-\bar{b}})}{f_{\rm BW}(m_t)}
\left|\tilde{\cal A}^{(t\bar{t})}\right|^2
\end{equation}
locally in phase space. The momenta are reorganized so as to put the $\bar{t}$
quark on its mass shell. Although gauge invariant, this procedure is still
somewhat arbitrary.
We therefore introduce here a third option, where nothing is removed or subtracted
from the associated production, but simply the full production cross section is
retained. Once a sample of events is generated, one can then still decide to
remove events near the resonance region and replace them with events obtained, for
example, with a full NLO implementation of $t\bar{t}$ production.

In our POWHEG code, we implemented all three methods described above. DR is the
simplest case. If the flag {\tt DR} is set to one in the file {\tt powheg.input},
the resonant diagrams of Fig.\ \ref{resdiag} are simply not included. For the
other two procedures, i.e.\ DS and keeping the full cross section, {\tt DR} should
be set to zero. The $s$-channel propagators of the $\bar{t}$ quark in the real
amplitudes are then replaced by a Breit-Wigner form. Setting the flag {\tt DS} to
one turns on diagram subtraction. If neither {\tt DS} nor {\tt DR} are set to one,
the full cross section is computed. In this case it is, however, hard to probe the
$\bar{t}$ pole with sufficient accuracy in the Monte Carlo integration. An
additional flag {\tt sepresonant} is therefore introduced that, when set to one,
causes POWHEG to treat the resonant contributions as a regular remnant. This is
possible since they do not require subtractions. A specific routine for the
generation of the phase space of the regular remnant ensures that appropriate
importance sampling is used in the $\bar{t}$ resonant region.

While with the DR or the full scheme the fraction of negative weights is very
small, this is not the case in the DS scheme. Here the real cross section can
become negative in certain kinematic regions, so that POWHEG must then be run with
the flag {\tt withnegweights} set to one. Negatively weighted events are then
kept, but are hard to interpret, since they correspond to the subtraction of an
ad hoc quantity from the cross section.

Removing diagrams at the amplitude level causes the loss of gauge invariance. A
considerable part of Ref.\ \cite{DRDS:2008} has been dedicated to the analysis
of the corresponding impact on $Wt$ production. There, different gauges were
considered for the gluon propagator, and differences at the per-mille level were
found.
%
Note, however, that gauge invariance is not only spoiled through the gluon
propagator, but also when the polarization sum
\begin{equation}
 P^{\mu \nu}(k)= \sum_{\lambda = 1,2} \epsilon^{\mu}(k,\lambda) \epsilon^{\nu}(k,\lambda)
\end{equation}
of external gluons is replaced by
\begin{equation}\label{simplepol}
 P^{\mu \nu}(k)= -g^{\mu \nu}
\end{equation}
for simplicity. Here, $k^\mu$ is the four-momentum, $\lambda$ is the polarization,
and $\epsilon^\mu(k,\lambda)$ is the polarization vector of the external gluon.
Eq.\ (\ref{simplepol}) includes not only physical transverse, but also
non-physical gluon polarizations that must be canceled by ghost contributions.
Removing individual diagrams then causes the loss of gauge invariance.
We therefore abandon the use of the simple polarization sum, Eq.\
(\ref{simplepol}), and sum instead only over physical states with
\begin{equation}
 P^{\mu \nu}(k) = -g^{\mu \nu} - \frac{1}{\left(k \cdot \eta \right)^2} \bigl[ \eta^2 k^{\mu} k^{\nu} - k \cdot \eta \left( k^{\mu} \eta^{\nu} + \eta^{\mu} k^{\nu}\right)  \bigr],
\end{equation}
where $\eta^\mu$ is an arbitrary four-vector transverse to the polarization vector
$\epsilon^\mu$.
When calculating a gauge invariant quantity, the $\eta$-dependence would drop out,
but this will not be the case in DR as argued above. For the channels with two
external gluons and incoming four-momenta $p_1$ and $p_2$, we choose for the
polarization vectors
\begin{eqnarray}
 \eta_1 ~=~ p_2 &{\rm and}& \eta_2 ~=~ p_1.
\end{eqnarray}

\section{Numerical results}
\label{sec:4}

\subsection{QCD input}

For the parton density functions (PDFs) in the external hadrons, we use the
set CT10 obtained in the latest global fit by the CTEQ collaboration
\cite{Lai:2010vv}. It has been performed at NLO in the $\overline{\rm MS}$
factorization scheme with $n_f=5$ active flavors as required by our calculation.
The employed value of $\alpha_s(M_Z)=0.118$, close to the world average, is
equivalent to setting the QCD scale parameter $\Lambda_{\overline{\rm MS}}^{n_f=
5}$ to 226.2 MeV as in the previous fits. We also adopt their value for the bottom
quark mass of $m_b=4.75$ GeV and for the top-quark mass of $m_t=172$ GeV and not
the newest average value of $m_t=173.2$ GeV obtained in direct top observation at
the Tevatron \cite{:2009ec}, as the former value corresponds nicely to the one in
the MC@NLO publication \cite{Weydert:2009vr} that we will compare with later in
this section. For the sake of easier comparisons we also adopt the default scale
choice $\mu_F=\mu_R=(m_H+m_t)/2$ as in the MC@NLO study.
We use the versions HERWIG 6.5.10 and PYTHIA 6.4.21 with stable top quarks and
Higgs bosons and no kinematic cuts to simplify the analysis. Multiparticle
interactions were neglected.
For a discussion of
the numerical impact of the bottom mass in the PDFs we refer the reader to Ref.\
\cite{Plehn:2010gp}. 

\subsection{Two Higgs Doublet Models}
New particles with masses in the TeV range, that couple to quarks at the tree
level, can strongly modify the predictions for Flavor Changing Neutral Current
(FCNC) processes, since these are absent at tree level in the SM. Thus all
extensions of the SM, including the 2HDMs, must avoid conflicts with the strict
limits on FCNCs, such as the electroweak precision observable $R_b=\Gamma(Z\to
b\bar{b})/\Gamma(Z\to{\rm hadrons})$ or the branching ratio BR($B\to X_s\gamma$).
In 2HDMs, tree-level FCNCs are traditionally avoided by imposing the hypothesis of
Natural Flavor Conservation (NFC), which allows only one Higgs field to couple to
a given quark species due to the presence of a flavor-blind Peccei-Quinn $U(1)$ symmetry
or its discrete subgroup $Z_2$ \cite{Glashow:1976nt}. Alternatively, all
flavor-violating couplings can be linked to the known structure of Yukawa
couplings and thus the Cabibbo-Kobayashi-Maskawa (CKM) matrix under the hypothesis
of Minimal Flavor Violation (MFV) \cite{D'Ambrosio:2002ex}. Both hypotheses have
recently been compared with the result that the latter appears to be more stable
under quantum corrections, but that the two hypotheses are largely equivalent at
tree level \cite{Buras:2010mh}. 

In a general 2HDM, one introduces two complex SU(2)-doublet scalar fields
\bea
 \Phi_i &=& \lr\begin{array}{c} \phi_i^+\\
 (v_i+\phi_i^{0,r}+i\phi^{0,i}_i)/\sqrt{2}\end{array}\rr ~~~{\rm with}~~~ i=1,2,
\eea
where the Vacuum Expectation Values (VEVs) $v_{1,2}$ of the two doublets are
constrained by the $W$-boson mass through $v^2=v_1^2+v_2^2=4m_W^2/g^2=(246~{\rm
GeV})^2$ \cite{Gunion:1989we}. The physical charged Higgs bosons are
superpositions of the charged degrees of freedom of the two doublets,
\bea
 H^\pm&=&-\sin\beta\,\phi_1^\pm + \cos\beta\,\phi_2^\pm,
\eea
and the tangent of the mixing angle $\tan\beta=v_2/v_1$, determined by the ratio
of the two VEVs, is a free parameter of the model, along with the mass of the
charged Higgs bosons $m_H$. The allowed range of $\tan\beta$ can be
constrained by the perturbativity of the bottom- and top-quark Yukawa couplings
($y_{t,b}\leq 1$) to $1 \leq \tan\beta \leq 41$. Note that in the Minimal
Supersymmetric SM (MSSM) $m_H>m_W$ at tree level. The possible assignments
of the Higgs doublet couplings to charged leptons, up- and down-type quarks
satisfying NFC are summarized in Tab.\ \ref{tab:1}.

\begin{table}
\caption{\label{tab:1}Couplings of the two Higgs doublets $\Phi_{1,2}$ to up-type
 quarks ($u$), down-type quarks ($d$), and charged leptons ($l$) in 2HDMs satisfying
 Natural Flavor Conservation \cite{Logan:2010ag}.}
\begin{tabular}{|l|cccc|}
\hline
Model    & Type-I & Lepton-specific & Type-II & Flipped \\
\hline
$\Phi_1$ & -      & $l$               & $d,l$     & $d$   \\
$\Phi_2$ & $u,d,l$  & $u,d$             & $u$       & $u,l$\\
\hline
\end{tabular}
\end{table}

In the Type-I 2HDM, only $\Phi_2$ couples to the fermions in exactly the same way
as in the minimal Higgs model, while $\Phi_1$ couples to the weak gauge bosons
\cite{Haber:1978jt}. The Feynman rules for the charged Higgs-boson couplings to
quarks in this model, with all particles incoming, are
\bea
 H^+\bar{u}_id_j &:& {ig\over\sqrt{2}M_W}V_{ij}(\cot\beta\,m_{u_i}P_L-\cot\beta\,
 m_{d_j}P_R),
\eea
where $V_{ij}$ is the CKM matrix and $P_{L,R}=(1\mp\gamma_5)/\sqrt{2}$ project out
left- and right handed quark eigenstates. As can be seen from Tab.\ \ref{tab:1},
these couplings are the same in the lepton-specific 2HDM.

In the Type-II 2HDM, $\Phi_2$ couples to up-type quarks and $\Phi_1$ to
down-type quarks and charged leptons. The Feynman rules for charged Higgs-boson
couplings to quarks in this model, with all particles incoming, are
\bea
 H^+\bar{u}_id_j &:& {ig\over\sqrt{2}M_W}V_{ij}(\cot\beta\,m_{u_i}P_L+\tan\beta\,
 m_{d_j}P_R).
\eea
As can again be seen from Tab.\ \ref{tab:1}, they are identical to those in the
flipped 2HDM \cite{Logan:2010ag}.

Since the NFC and MFV hypotheses allow for the possibility that the two Higgs
doublets couple to quarks with arbitrary coefficients $A_{u,d}^i$, there exists also
the possibility of a more general 2HDM, sometimes called Type-III 2HDM
\cite{Degrassi:2010ne}. In this case, the Feynman rules for the charged Higgs-boson
couplings to quarks, with all particles incoming, are
\bea
 H^+\bar{u}_id_j &:& {ig\over\sqrt{2}M_W}V_{ij}(A_u^i\,m_{u_i}P_L-A_d^i\,
 m_{d_j}P_R),
\eea
where the family-dependent couplings $A_{u,d}^i$ read
\bea
 A_{u,d}^i &=& A_{u,d}\lr1+\eps_{u,d}{m_t^2\over v^2}\delta_{i3}\rr.
\eea
Under the assumption that no new sources of CP violation apart from the
complex phase in the CKM matrix are present, the coefficients $A_{u,d}$
and $\eps_{u,d}$ are real. The case $\eps_{u,d}=0$ corresponds to the NFC
situation, in which the Yukawa matrices of both Higgs doublets are aligned
in flavor space.
LEP measurements of $R_b$ constrain $|A_u|$ to values below 0.3
and 0.5 (0.78 and 1.35) for $m_H=100$ and 400 GeV at 1$\sigma$ (2$\sigma$),
when $A_d=0$. For opposite (same) signs of $A_u$ and $A_d$, the average of BABAR,
Belle and CLEO measurements of BR($B\to X_s\gamma$) allow for one (two) region(s)
of $A_d$ for given values of $A_u$ and $m_H$. For $A_u=0.3$ and
$m_H=100$ GeV, both $A_d\in[0;1]$ and [16;18] are allowed, while for
$A_u=0.3$ and $m_H=400$ GeV, both $A_d\in[0;2.5]$ and [50;56] are allowed
at 2$\sigma$ \cite{Degrassi:2010ne}.
Since general color-singlet Higgs-boson couplings and (theoretically possible) color-octet
Higgs bosons induce different QCD corrections, we will not study these
scenarios numerically.
In the literature, one may also find Type-III
2HDMs where no flavor symmetry is imposed and FCNCs are avoided by other
methods, e.g.\ by the small mass of first and second generation quarks
\cite{Cheng:1987rs}. These models then allow for the couplings of charged Higgs
bosons to bottom and charm quarks, which induces a phenomenology that is different
from the one studied in this paper.

\subsection{Predictions for various 2HDMs}

The calculation presented in the previous sections was performed in a generic way, which
makes it possible to use the result for various models with charged Higgs bosons.
Out of the models mentioned in the last section, our calculation is in particular
valid for the Type-I and Type-II 2HDMs. In this subsection, we perform a numerical
analysis for a set of typical collider scenarios for these 2HDMs.
As in these models the scattering matrix element is directly proportional to the Higgs-top-bottom
quark coupling even at NLO, the type of the model has no influence on kinematic
distributions apart from their normalization to the total cross section.

We therefore concentrate here on the total cross sections and on the uncertainties
both from the variation of the renormalization and factorization scales and from the
parton distribution functions. For a better comparison, we analyze the total cross
sections and their uncertainties by choosing the same values for the mass of the
charged Higgs boson and for $\tan\beta$ in all scenarios, i.e. $m_H=300$ GeV and
$\tan\beta = 10$. All relevant values are summarized in Tab.\ \ref{totCS:tab}.	
\begin{table}
\caption{Total cross sections (in pb) for different 2HDMs at the Tevatron and at the LHC at
 leading order (LO) and at next-to-leading order (NLO) including the scale and PDF
 uncertainties. All scenarios assume the same parameters for better comparison, i.e.\
 $m_H=300$ GeV and $\tan\beta = 10$.}\label{totCS:tab}
\begin{tabular}{|c|c|c|c|c|c|}
\hline
Scenario & LO & Scale unc. & NLO & Scale unc. & PDF error \\
\hline
Tevatron 2HDM-I  & $3.229.10^{-6}$ & $\left.\right.^{+1.306.10^{-6} (40\%)}_{-0.901.10^{-6} (28\%)}$ 
		         & $6.218.10^{-6}$ & $\left.\right.^{+1.388.10^{-6} (22\%)}_{-1.201.10^{-6} (19\%)}$ 
								   & $\left.\right.^{+4.448.10^{-5} (72\%)}_{-2.362.10^{-5} (38\%)}$ \\
Tevatron 2HDM-II & $1.303.10^{-5}$ & $\left.\right.^{+0.524.10^{-5} (40\%)}_{-0.365.10^{-5} (28\%)}$ 
                 & $2.506.10^{-5}$ & $\left.\right.^{+0.565.10^{-5} (23\%)}_{-0.484.10^{-5} (19\%)}$ 
                                   & $\left.\right.^{+1.792.10^{-5} (72\%)}_{-0.952.10^{-5} (38\%)}$ \\
\hline 	
LHC 2HDM-I       & $1.577.10^{-3}$ & $\left.\right.^{+0.379.10^{-3} (24\%)}_{-0.304.10^{-3} (19\%)}$ 
		         & $2.189.10^{-3}$ & $\left.\right.^{+0.162.10^{-3} (7\%)}_{-0.199.10^{-3} (9\%)}$ 
  					               & $\left.\right.^{+0.356.10^{-3} (16\%)}_{-0.304.10^{-3} (14\%)}$ \\
LHC 2HDM-II      & $6.366.10^{-3}$ & $\left.\right.^{+1.514.10^{-3} (24\%)}_{-1.237.10^{-3} (19\%)}$ 
		         & $8.821.10^{-3}$ & $\left.\right.^{+0.651.10^{-3} (7\%)}_{-0.802.10^{-3} (9\%)}$ 
  					               & $\left.\right.^{+1.433.10^{-3} (16\%)}_{-1.223.10^{-3} (14\%)}$ \\
\hline 	
\end{tabular}
\end{table}

In all scenarios, both at the Tevatron and at the LHC, the next-to-leading order correction is
substantial, ranging from $57\%$ at the Tevatron in the Type-I 2HDM to $38\%$ at the LHC in the
same model. Apart from enhancing the total cross section, including the NLO correction reduces
the theoretical error defined as the scale uncertainty of the cross section. The scale
uncertainty is obtained by varying both the renormalization and factorization scales
simultaneously in the interval
\begin{equation}
	\frac{m_t+m_H}{4} < \mu < m_t+m_H\,.
\end{equation} 
At leading order, the strong scale dependence comes from the strong coupling constant and from
the Yukawa coupling in the tree-level amplitude. Including higher-order corrections, this
uncertainty is dramatically reduced in some scenarios.

Another large source of error stems from the parton distribution functions. We use the CT10 NLO
PDF set with its error PDF sets to determine the error coming from the uncertainty contained in
determining the parton content of the colliding hadrons. The process considered here is extremely
sensitive to the gluon distribution function through having a gluon in the initial state
and through having a heavy-quark initial state, which is radiatively generated from the gluon
PDF. Moreover, the production of a heavy Higgs boson in association with a top quark probes the
higher $x$ content of the initial-state (anti-)proton. The values of Bjorken-$x$ probed can be
expressed as
\begin{equation}
	x_a x_b = \frac{(k_1+k_2)^2}{s} > \frac{(m_t+m_H)^2}{s},
\end{equation} 
which at the Tevatron leads to typical values of $x\sim 0.3$. This is exactly the region where
the gluon PDF is poorly known, which translates into large PDF uncertainties on the cross
section at the Tevatron. At the LHC, the Bjorken-$x$ probed is $x\sim 0.1$, and 
the PDF uncertainties are therefore much smaller.

\subsection{Checks of the NLO calculation and comparisons with POWHEG}

As a check of the numerical implementation of our analytical results, we have
compared our complete NLO calculation obtained with the Catani-Seymour dipole formalism
with the one performed previously with a phase-space slicing method using a single
invariant-mass cutoff \cite{Plehn:2002vy}, which had in turn been found to agree
with a calculation using a two (soft and collinear) cutoff phase-space slicing method
\cite{Zhu:2001nt}. We found good agreement for all differential and total cross sections
studied, but refrain from showing the corresponding figures here, since the fixed-order
results are well-known.

\begin{figure}
 \centering
 \epsfig{file=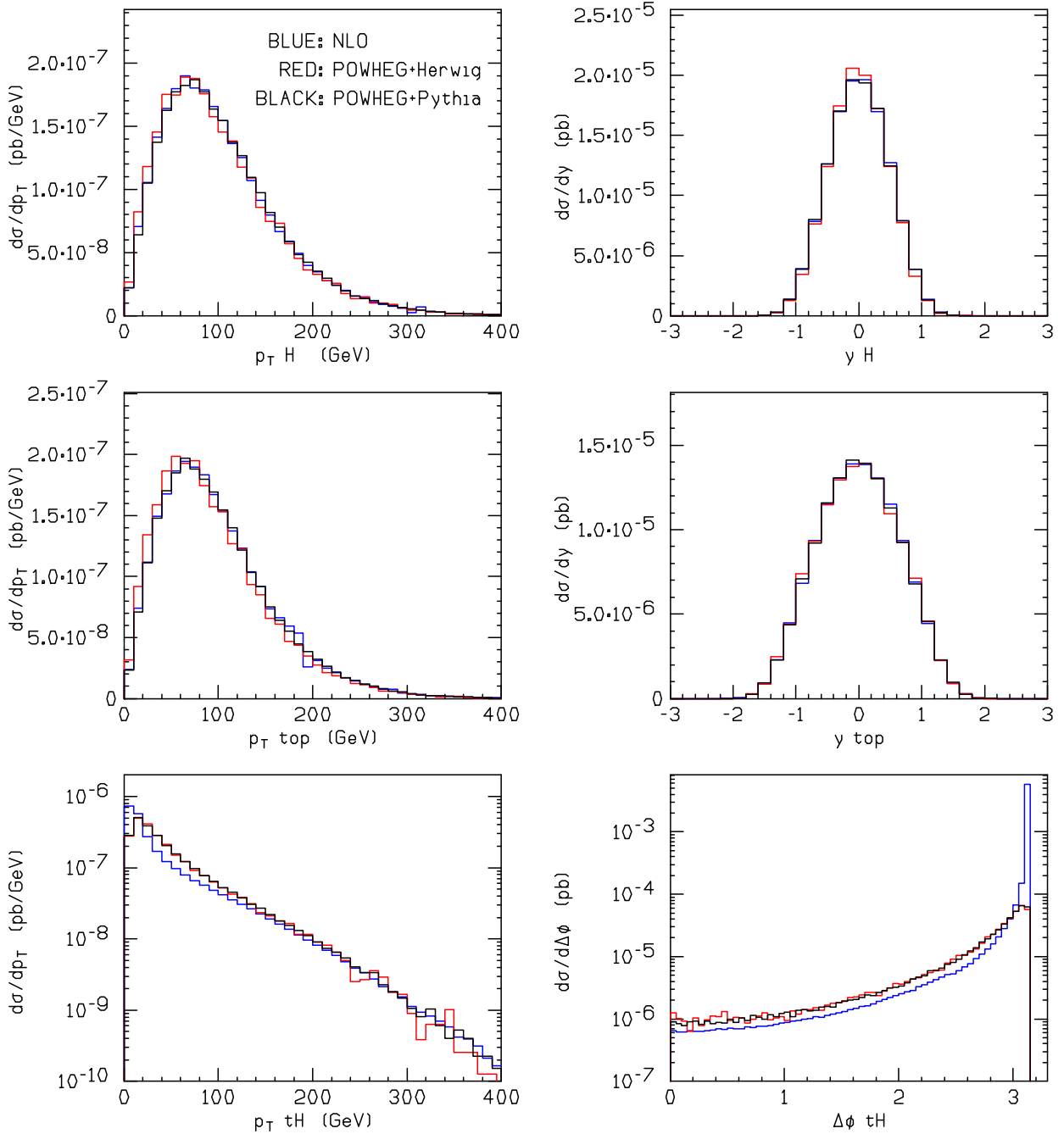,angle=90,width=\textwidth}
 \caption{\label{fig:4}Distributions in transverse momentum $p_T$ (top left) and
 rapidity $y$ (top right) of the charged Higgs boson, $p_T$ (center left) and $y$
 (center right) of the top quark, as well as $p_T$ (bottom left) and azimuthal opening angle
 $\Delta\phi$ (bottom right) of the $tH^-$ system produced at the Tevatron with $\sqrt{S}
 =1.96$ TeV. We compare the NLO predictions without (blue) and with matching to
 the PYTHIA (black) and HERWIG (red) parton showers using POWHEG in the Type-II
 2HDM with $\tan\beta=10$ and $m_H=300$ GeV.}
\end{figure}

\begin{figure}[!h]
 \centering
 \epsfig{file=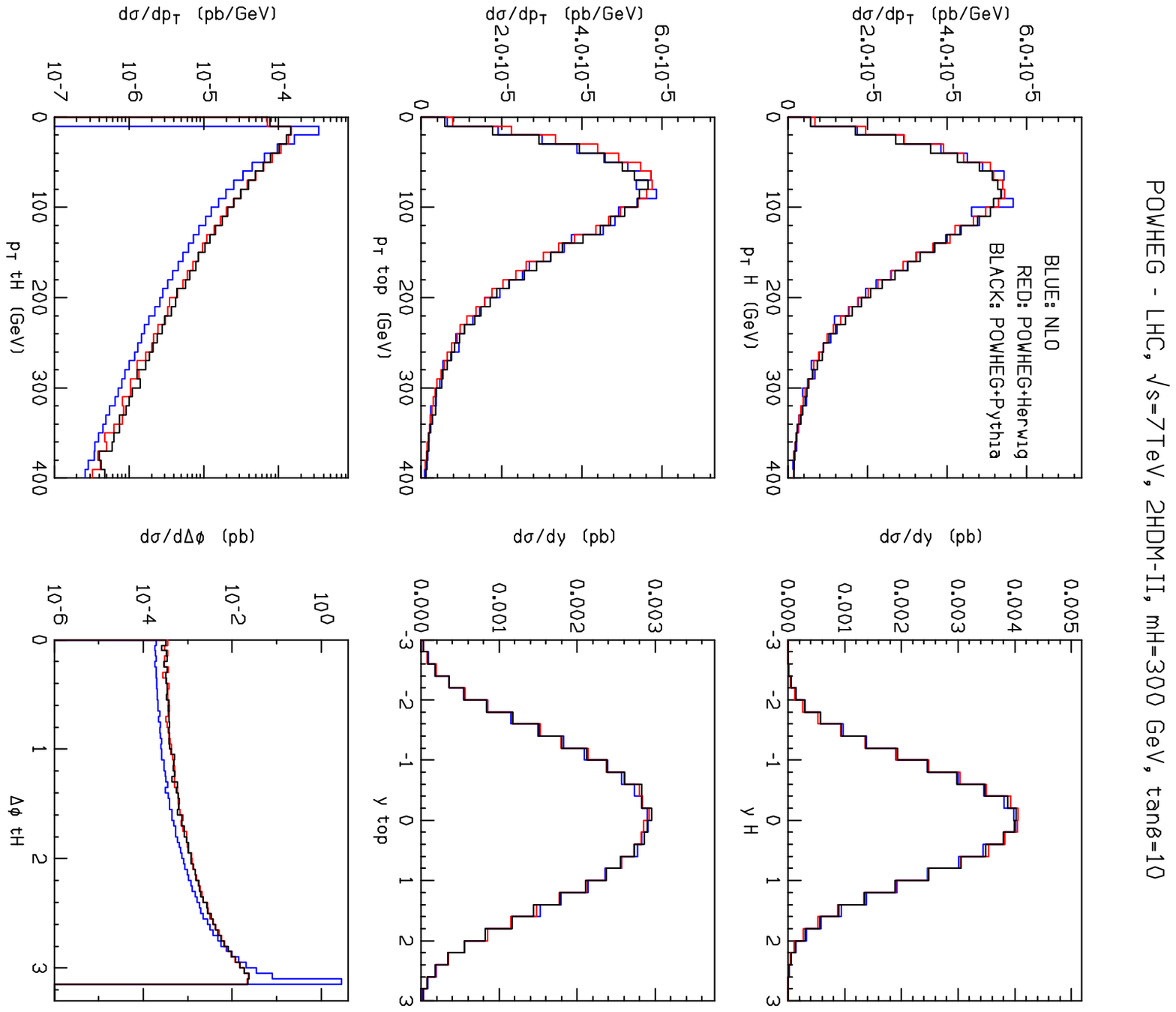,angle=90,width=\textwidth}
 \caption{\label{fig:5}Same as Fig.\ \ref{fig:4} at the LHC with $\sqrt{S}=7$
 TeV.}
\end{figure}

\begin{figure}[!h]
 \centering
 \epsfig{file=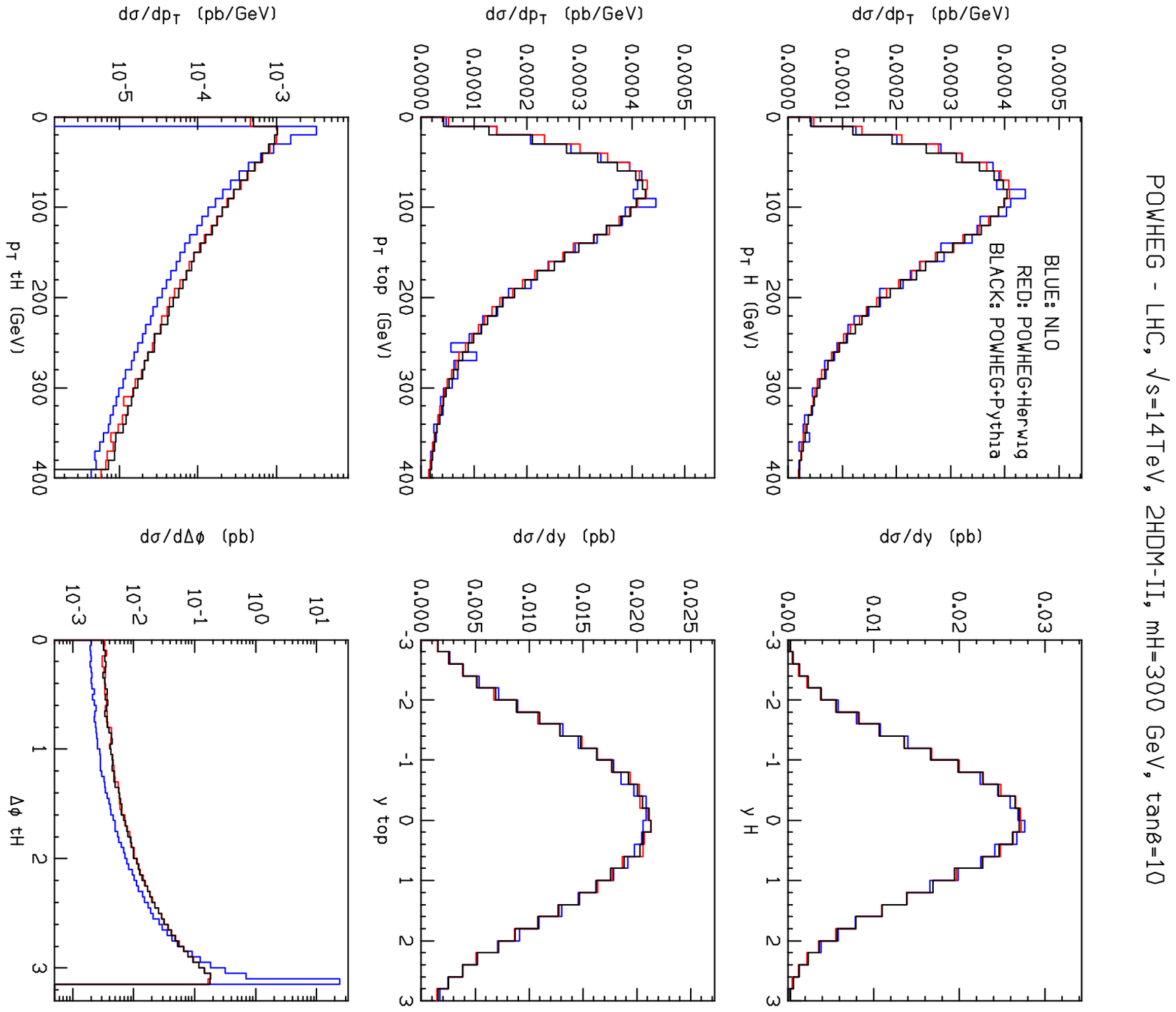,angle=90,width=\textwidth}
 \caption{\label{fig:6}Same as Fig.\ \ref{fig:4} at the LHC with $\sqrt{S}=14$
 TeV.}
\end{figure}

For the remainder of the analysis, we will constrain ourselves to the Type-II 2HDM,
as the kinematic distributions have the same features in both Type-I and Type-II 2HDMs.
In all of our discussion, we consider three collider scenarios:
\begin{itemize}
	\item Tevatron, $\sqrt{S}=1.96\ {\rm TeV}$,
	\item LHC, $\sqrt{S}=7\ {\rm TeV}$, and
	\item LHC, $\sqrt{S}=14\ {\rm TeV}$.
\end{itemize}
Moreover, in the comparison of our NLO calculation with our implementation of its relevant 
parts in the POWHEG BOX, we assume $m_H=300$ GeV and $\tan\beta=10$. The results are shown in
Fig.\ \ref{fig:4} for the Tevatron with $\sqrt{S}=1.96$ TeV and Figs.\ \ref{fig:5} and
\ref{fig:6} for the LHC with a center-of-mass energy of $\sqrt{S}=7$ and 14 TeV, respectively.

%
%
%
%

%
%
%
\begin{figure}
 \centering
 \epsfig{file=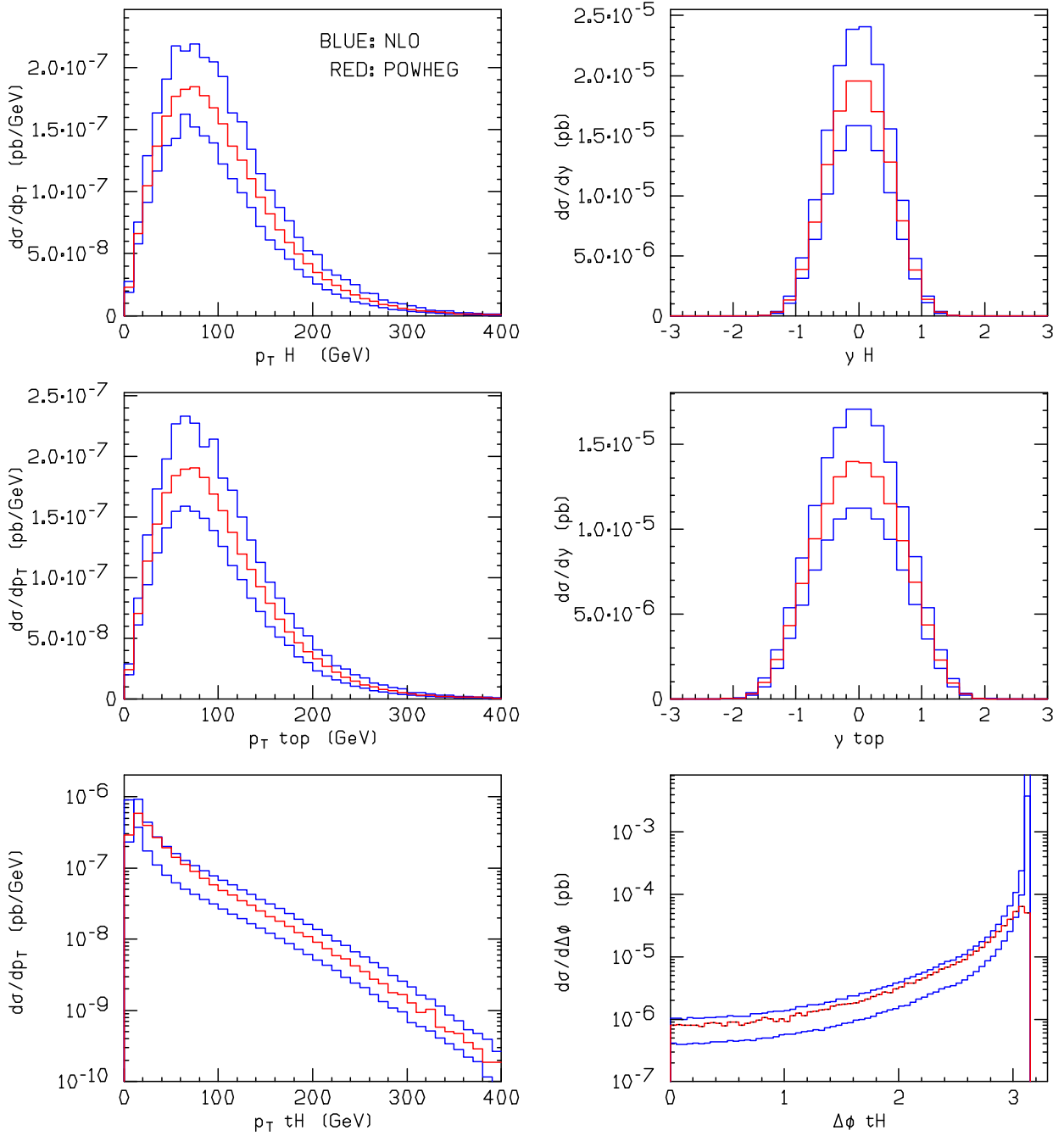,angle=90,width=\textwidth}
 \caption{\label{fig:4a}Distributions in transverse momentum $p_T$ (top left) and
 rapidity $y$ (top right) of the charged Higgs boson, $p_T$ (center left) and $y$
 (center right) of the top quark, as well as $p_T$ (bottom left) and azimuthal opening angle
 $\Delta\phi$ (bottom right) of the $tH^-$ system produced at the Tevatron with $\sqrt{S}
 =1.96$ TeV. We compare the NLO scale uncertainty band (blue) the POWHEG result
including first radiation only (red).}
\end{figure}
%
%
\begin{figure}[!h]
 \centering
 \epsfig{file=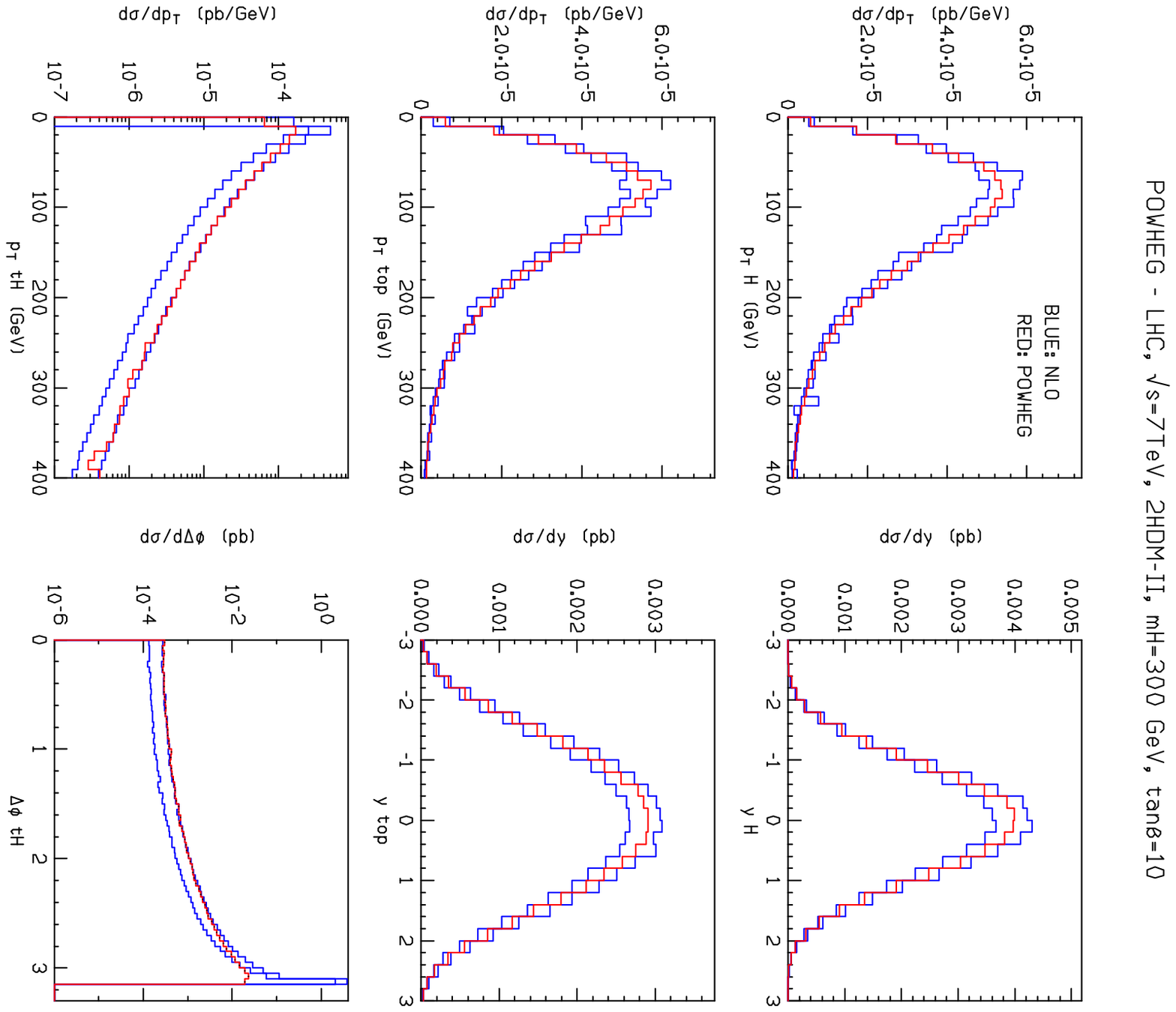,angle=90,width=\textwidth}
 \caption{\label{fig:5a}Same as Fig.\ \ref{fig:4a} at the LHC with $\sqrt{S}=7$
 TeV.}
\end{figure}
%

%
\begin{figure}[!h]
 \centering
 \epsfig{file=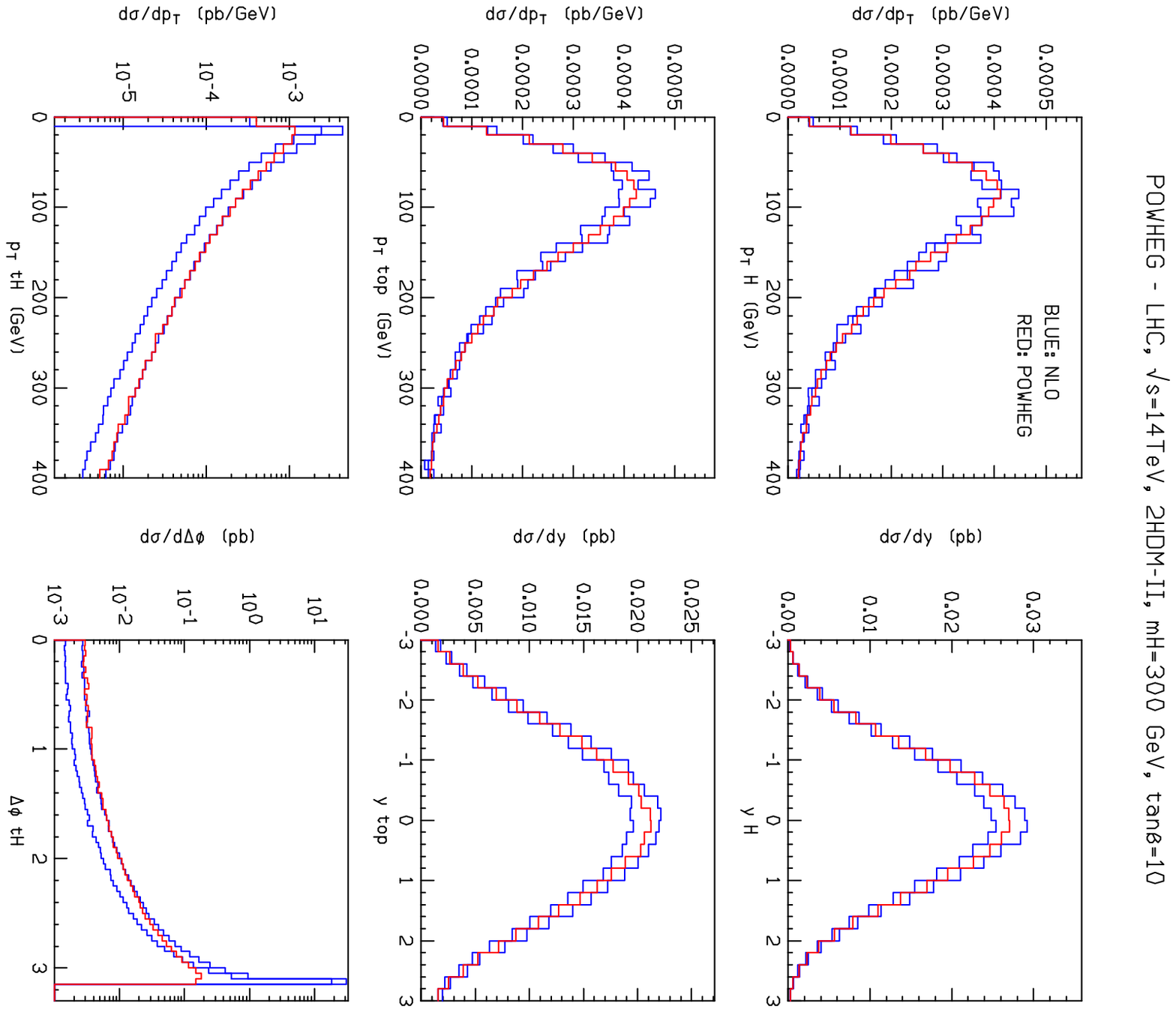,angle=90,width=\textwidth}
 \caption{\label{fig:6a}Same as Fig.\ \ref{fig:4a} at the LHC with $\sqrt{S}=14$
 TeV.}
\end{figure}
%

If we concentrate first on the transverse-momentum ($p_T$, left) and rapidity ($y$, right)
distributions of the charged Higgs boson (top) and top quark (center) individually, we
observe good agreement in absolute normalization and shape for all three collider scenarios,
independently if a parton shower is matched to the NLO calculation or not. This corresponds
to the well-known fact that these distributions are largely insensitive to soft or
collinear radiation, in particular from the initial state, and this can therefore be seen as
a further consistency test of our calculations.
Soft radiation becomes relevant in all three collider scenarios when we consider the azimuthal
opening angle of the top-Higgs pair (bottom right), where the singularity occurring at
NLO in back-to-back kinematics at  $\Delta\phi=\pi$ is regularized and resummed by the
parton showers. This holds also for the $p_T$-distribution of the top-Higgs pair
(bottom left), which diverges perturbatively at $p_T=0$ GeV and even turns negative
at the LHC.

An advantage of the POWHEG method is that it can also provide events
including first radiation only in the form of an event file according to the Les
Houches format (LHEF), making them independent of the parton shower. We therefore
compare in Figs.\ \ref{fig:4a}--\ref{fig:6a} the distributions obtained from these
files to those obtained with NLO accuracy for the same set of parameters as in
Figs.\ \ref{fig:4}--\ref{fig:6}. As one can clearly see, they lie within the NLO
scale uncertainty band, showing that the difference comes from terms beyond
NLO accuracy. This provides a good consistency check of the matching procedure.

\subsection{POWHEG predictions with HERWIG and PYTHIA parton showers}

In Figs.\ \ref{fig:4}--\ref{fig:6}, we also show two different predictions with POWHEG
coupled either to the angularly-ordered HERWIG or to the virtuality-ordered PYTHIA
parton shower. The agreement of the HERWIG and PYTHIA results is in general very good.
They differ only slightly in the $p_T$ distributions of the top-Higgs
pair, where the PYTHIA $p_T$-distribution is a little bit harder, in particular at the
Tevatron.

\subsection{POWHEG comparison with MATCHIG}

As described in Sec.\ \ref{sec:2}, the production of charged Higgs bosons and top quarks
proceeds at LO through the process $bg\to H^-t$, while at NLO the process $gg\to H^-t\bar{b}$
appears. The latter implies the creation of a virtual initial $b$-quark,
which may either occur in the perturbative part of the calculation or is resummed into a
$b$-quark PDF. In the full NLO calculation, the separation is achieved through the factorization
procedure and induces a dependence on the factorization scale $\mu_F$.

Before schemes to match parton showers with full NLO calculations were developed, the importance
of the contribution of this particular two-to-three process and the perturbative origin of the
$b$-quark density had already been recognized \cite{Alwall:2005gs}. It had been proposed to
supplement the LO calculation by this particular two-to-three process and to remove the overlap
by subtracting the doubly counted (DC) term
\bea
 \sigma_{\rm DC} &=& \int_0^1 dx_a\, f_{b'}(x_a,\mu_F^2) \int_0^1 dx_b \, f_{g}(x_b,\mu_F^2)
 \sigma^{LO}(p_1,p_2) + (x_a\leftrightarrow x_b),
\eea
where $f_{b'}(x,\mu_F^2)$ is the LO $b$-quark density given by
\bea
 f_{b'}(x,\mu_F^2)&\simeq&{\alpha_s\over2\pi}\ln{\mu_F^2\over m_b^2}
 \int {dz\over z} P_{qg}(z) f_{g}\lr {x\over z},\mu_F^2\rr
\eea
with $P_{qg}(z)$ the $g\to q$ splitting function, $f_g(x,\mu_F^2)$ the gluon PDF,
and $z$ the longitudinal gluon momentum fraction taken by the $b$-quark.
The two-to-three and double-counting processes had been implemented in an
addition to PYTHIA called MATCHIG.
%
\begin{figure}[!p]
 \centering
 \epsfig{file=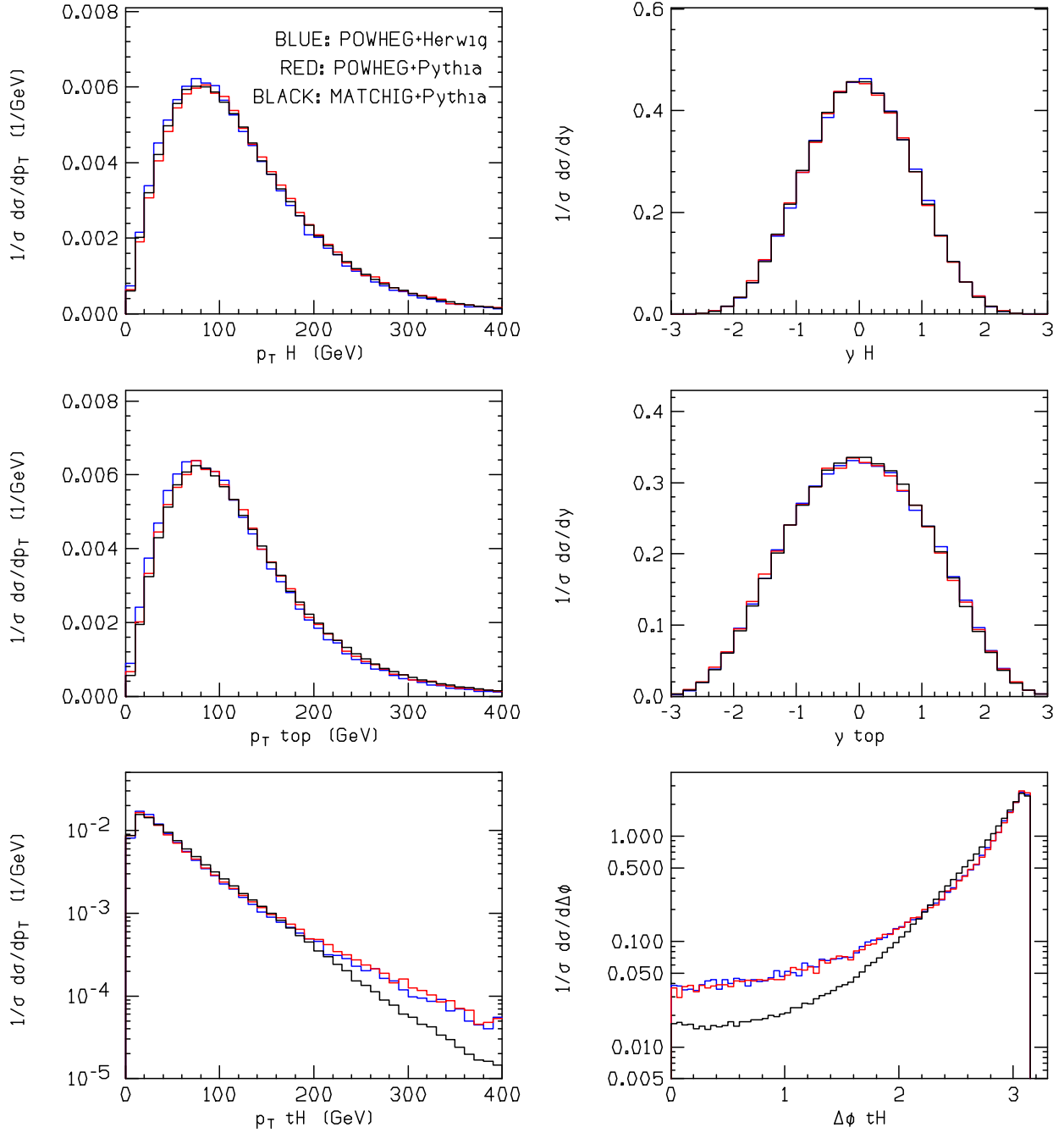,angle=90,width=\textwidth}
 \caption{\label{fig:11}Distributions in transverse momentum $p_T$ (top left) and
 rapidity $y$ (top right) of the charged Higgs boson, $p_T$ (center left) and $y$
 (center right) of the top quark, as well as $p_T$ (bottom left) and azimuthal opening angle
 $\Delta\phi$ (bottom right) of the $tH^-$ system produced at the LHC with $\sqrt{S}
 =7$ TeV. We compare the tree-level predictions matched to PYTHIA using
 MATCHIG (black) with our NLO calculation matched to PYTHIA (red) and HERWIG (blue)
 using POWHEG. 
 All distributions have been normalized to the respective total cross sections.}
\end{figure}
%

With our full NLO calculation matched to PYTHIA within the POWHEG BOX, it is now
possible to compare the two approaches numerically. The results are shown in
Fig.\ \ref{fig:11}. Since the normalization of the MATCHIG prediction is still
effectively of LO, we have normalized all distributions to their respective total
cross sections in order to emphasize the shapes of the distributions. One observes
that when both the MATCHIG (black) and POWHEG (red) predictions are matched to the
PYTHIA parton shower, there is very little difference, even at low $p_T$ and large
$\Delta\phi$ of the top-Higgs pair. Only at large $p_T$ and small
$\Delta\phi$ the differences become sizable, which can be attributed to the fact
that MATCHIG includes only one of the four classes of real-emission processes,
while our POWHEG prediction includes also the quark-initiated real-emission
processes. Let us emphasize again that while the spectra are already quite well
described with MATCHIG, their normalization is only accurate to LO and not NLO as
in POWHEG.

\subsection{Comparison with MC@NLO}

\begin{figure}
 \centering
 \epsfig{file=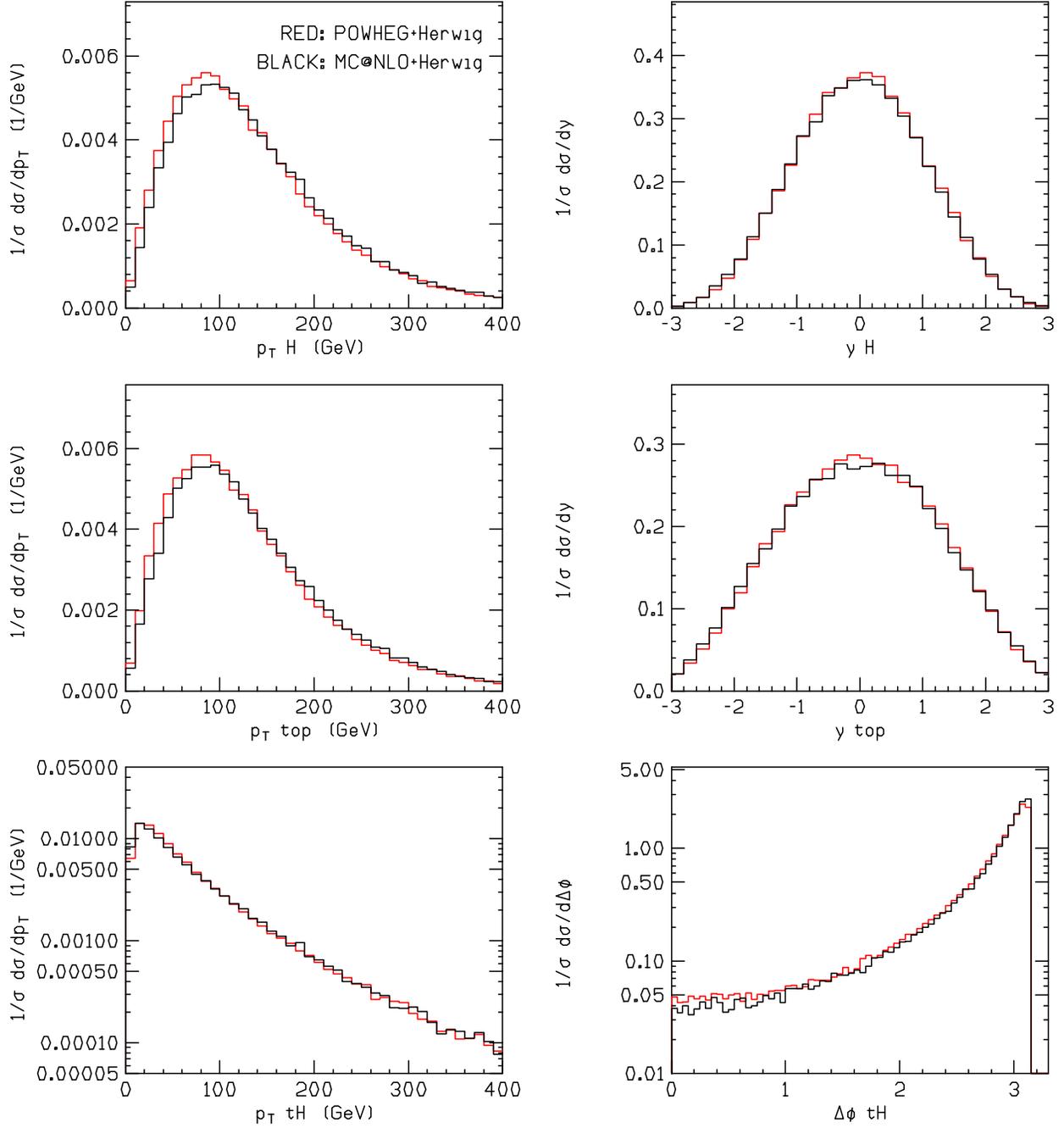,angle=90,width=\textwidth}
 \caption{\label{fig:7}Distributions in transverse momentum $p_T$ (top left) and
 rapidity $y$ (top right) of the charged Higgs boson, $p_T$ (center left) and $y$
 (center right) of the top quark, as well as $p_T$ (bottom left) and azimuthal opening angle
 $\Delta\phi$ (bottom right) of the $tH^-$ system produced at the LHC with $\sqrt{S}
 =14$ TeV. We compare the NLO predictions with matching to the HERWIG parton
 showers using POWHEG (red) and MC@NLO (black) in the Type-II 2HDM with $\tan
 \beta=30$ and $m_H=300$ GeV. All distributions have been normalized to the
 respective total cross sections.}
\end{figure}

In a recent publication, two of us and a number of other authors have matched a NLO
calculation performed with the FKS subtraction formalism to the HERWIG PS with the
MC@NLO method \cite{Weydert:2009vr}. It is therefore mandatory that we compare in
this paper this previous work with our new POWHEG implementation, which we do in
Fig.\ \ref{fig:7}. Note that here we employ a value of $\tan\beta=30$ as in the
%
%
MC@NLO publication. In both calculations, we use the HERWIG PS in order to emphasize
possible differences in the matching methods and not those in the parton shower.
We also normalize the differential cross sections again to the total cross section for a better
comparison of the shapes of the distributions.

As in the other comparisons, the rapidity distributions of the charged Higgs boson
(top right) and the top quark (center right) show little variation, confirming
the consistency of the two calculations. However, the corresponding $p_T$-spectra
(top and center left) are slightly harder with the MC@NLO matching than in POWHEG.
%
This behaviour is known from other processes \cite{Re:2010bp,Alioli:2009je}.
%
It is less pronounced in the $p_T$-distribution
of the top-Higgs pair, shown on a logarithmic scale (bottom left). Since we are using
the HERWIG PS, the rise at small azimuthal angle $\Delta\phi$ (bottom right)
is not very strong with MC@NLO and only slightly more so with POWHEG.
In total, all of these differences are similarly small in
the production of a top quark with a $W$-boson \cite{Re:2010bp} and with a charged
Higgs boson at the LHC.

\subsection{Diagram Removal, Diagram Subtraction, and no subtraction}

\begin{figure}
 \centering
 \epsfig{file=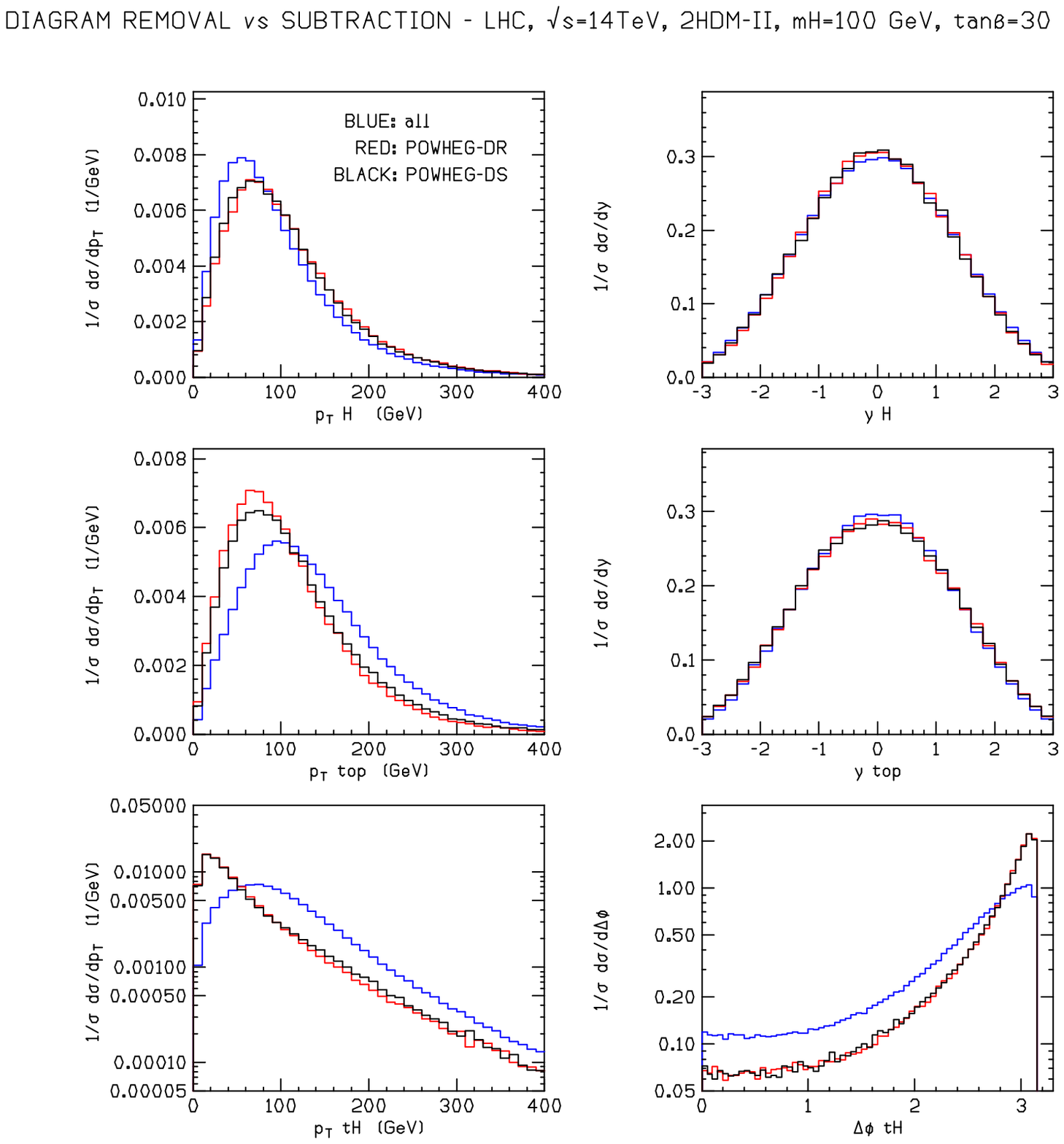,angle=90,width=\textwidth}
 \caption{\label{fig:10}Distributions in transverse momentum $p_T$ (top left) and
 rapidity $y$ (top right) of the charged Higgs boson, $p_T$ (center left) and $y$
 (center right) of the top quark, as well as $p_T$ (bottom left) and azimuthal opening angle
 $\Delta\phi$ (bottom right) of the $tH^-$ system produced at the LHC with $\sqrt{S}
 =14$ TeV. We compare the NLO predictions matched to the HERWIG parton shower using POWHEG
 with Diagram Removal (red), Diagram Subtraction (black), and without
 removing or subtracting anything (blue) in the Type-II 2HDM with $\tan\beta=30$ and
 $m_H=100$ GeV. All distributions have been normalized to the
 respective total cross sections.}
\end{figure}

If the charged Higgs boson was lighter than the top quark, it would dominantly be created
in top-pair production and the decay of an (anti-)top quark into it. As discussed above,
one must then find a suitable definition to separate this process from the associated
top-Higgs production discussed in this paper. In addition to the Diagram Removal (DR)
and Diagram Subtraction (DS) methods discussed above, we introduce here also the option
of not removing or subtracting anything from the associated production, but simply retaining
the total production cross section, which then allows for the removal of fully simulated events
near the resonance region and replacing them with events obtained, e.g., with a full
NLO implementation of $t\bar{t}$ production. The results are shown in Fig.\ \ref{fig:10}.
%
%
%
\begin{figure}
 \centering
 \epsfig{file=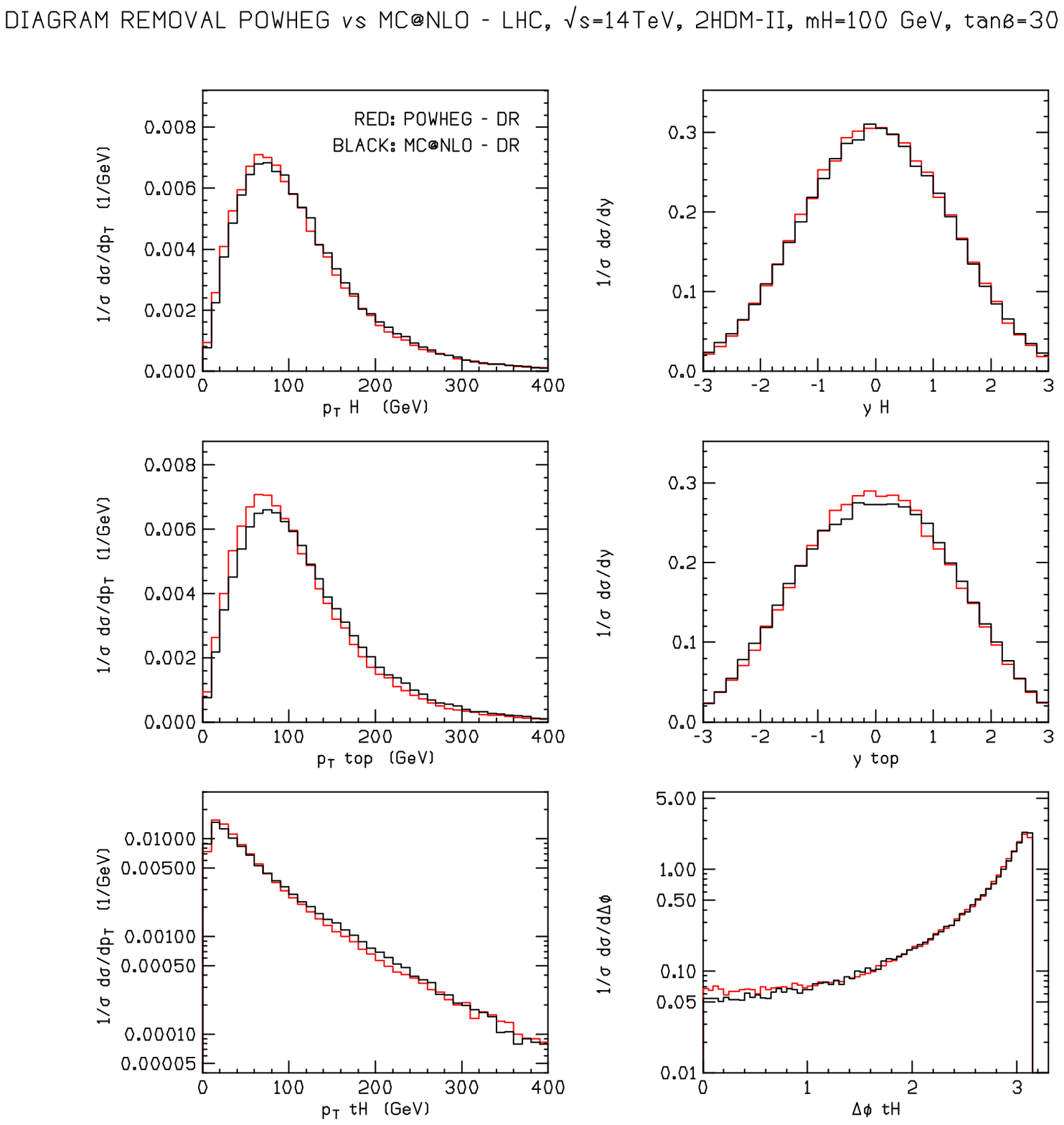,angle=90,width=\textwidth}
 \caption{\label{fig:8}Same as Fig.\ \ref{fig:7}, but for a light charged Higgs boson of mass
 $m_H=100$ GeV and using the DR method.}
\end{figure}
%
%
\begin{figure}
 \centering
 \epsfig{file=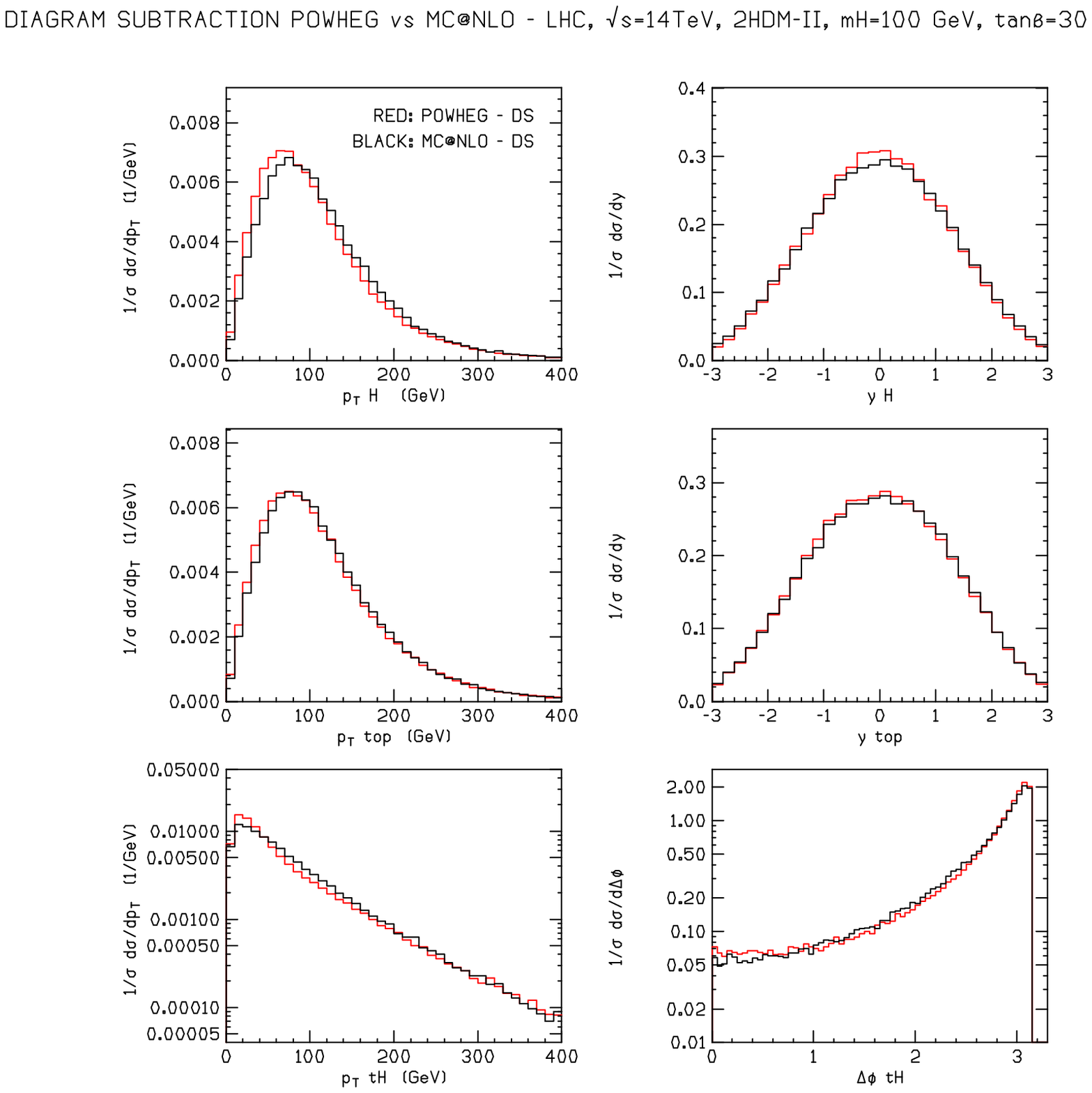,angle=90,width=\textwidth}
 \caption{\label{fig:9}Same as Fig.\ \ref{fig:7}, but for a light charged Higgs boson of mass
 $m_H=100$ GeV and using the DS method.}
\end{figure}
%

The rapidity distributions of the charged Higgs boson (top right) and top quark (center
right) show again little sensitivity to the different theoretical approaches. However, the
$p_T$-distribution of the charged Higgs boson (top left) is somewhat softer and the one of the
top quark (center left) considerably harder without removal or subtraction, as the difference
describes the distributions of the lighter decay product and the heavier decaying particle,
respectively. The $p_T$-distribution of the top-Higgs pair (bottom left) is significantly
harder (note again the logarithmic scale) and its maximum moves from $p_T=$ 20 to 70 GeV,
indicating that the transverse momentum of the pair is balanced by a hard object, i.e.\ the
fast additional $b$-quark jet, in the other hemisphere. This also allows the top-Higgs
pair to move closer together in azimuthal angle (bottom right).

The theoretical pros and cons and the numerical differences of Diagram Removal and Diagram
Subtraction have been discussed extensively above and also elsewhere \cite{Weydert:2009vr}.
It is clear from Fig.\ \ref{fig:10} that the numerical difference of DR vs.\ DS is much less
pronounced than the difference of both with respect to no removal or subtraction at all. We
emphasize that the total cross section is continuous across the $m_H=m_t$
threshold in all three schemes (see also Ref.\ \cite{Plehn:2010gp}).

The differences of POWHEG and MC@NLO are small for $m_H<m_t$ in both the
DR and DS schemes, as can be seen when comparing Figs.\ \ref{fig:8} and
\ref{fig:9}. This coincides nicely with our observation above
that these differences should be as small as in the associated production of
$W$-bosons and top quarks \cite{Re:2010bp}. 

\section{Conclusion}
\label{sec:5}

In this paper, we presented a new NLO calculation of the associated
production of charged Higgs bosons and top quarks at hadron colliders
using the Catani-Seymour dipole subtraction formalism and matched it
to parton showers with the POWHEG method. We discussed the different types of
2HDMs as well as the corresponding current experimental constraints and provided,
for specific benchmark values of the charged Higgs-boson mass and the ratio of
the two Higgs VEVs $\tan\beta$, the central values, scale, and PDF uncertainties
of the total cross sections at the Tevatron and LHC in tabular form for future
reference. As expected, the scale uncertainty was considerably reduced from
up to $\pm 100$\% at LO to less than $\pm 15$\% at NLO. However, the PDF uncertainty,
estimated with the CT10 set of global analyses, remained quite substantial, in
particular at the Tevatron, where high momentum fractions of the gluons and
$b$-quarks in the protons and antiprotons are probed.

For the differential cross sections, we established good numerical agreement
of our full NLO calculation with previous calculations. We then performed
detailed comparisons of our new POWHEG implementation with the purely
perturbative result, with PYTHIA or HERWIG parton showers, with a LO
calculation matched to the PYTHIA parton shower using MATCHIG, and with
a NLO calculation matched to the HERWIG parton shower using MC@NLO.

While the transverse-momentum distributions and the relatively central rapidity
distributions of the charged Higgs boson and top quark individually showed little
sensitivity to the existence and type of parton showers, the transverse-momentum
distribution of the top-Higgs pair depended quite strongly on the different
theoretical approaches as expected. This was also
true for the distribution in the azimuthal angle of the top-Higgs pair. For
scenarios in which the charged Higgs boson is lighter than the top quark,
we implemented in POWHEG in addition to the previously proposed Diagram
Removal and Diagram Subtraction schemes the possibility to retain the
full cross section and replace the simulated events in the resonance region
with a full NLO Monte Carlo for top-quark pair production.

It will now be very interesting to observe the impact of our work on the
experimental search for charged Higgs bosons. The numerical code and
technical support is, of course, available from the authors. 

\acknowledgments

This work has been supported by the French ANR through grants No.\
ANR-06-JCJC-0038-01 and ANR-07-BLAN-0245, by a Ph.D.\ fellowship of the French
Ministry for Education and Research, and by the Theory-LHC-France initiative of
the CNRS/IN2P3.


\end{document}

%% file: def.tex
\newcommand{\colm}{\mathbf{T}}
\newcommand{\ee}{\epsilon}
\def\Re  {{\rm Re}}
\def\Im  {{\rm Im}}
\def\KFS#1{K^{{#1}}_{\scriptscriptstyle\rm F\!.S\!.}}
\def\Ret {\widetilde{{\rm Re}}}
\def\veps{\varepsilon}
\def\hc    {{\rm h.c.}}
\newcommand{\fact}{{\mathrm{C}}}
\newcommand{\figref}[1]{Fig.~\ref{#1}}
\newcommand{\tabref}[1]{Tab.~\ref{#1}}
\newcommand{\rA}{{\mathrm{A}}}
\newcommand{\rB}{{\mathrm{B}}}
\newcommand{\rR}{{\mathrm{R}}}
\newcommand{\rV}{{\mathrm{V}}}
\def\bom#1{{\mbox{\boldmath $#1$}}}
\def\ket#1{|{#1}\rangle}
\def\bra#1{\langle{#1}|}
\def\sd{\tilde{d}}
\def\sq{\tilde{q}}
\def\su{\tilde{u}}
\def\mq{m_q}
\def\mqi{m_{q}}
\def\mqj{m_{q}}
\def\mqk{m_{q}}
\def\mql{m_{q}}
\def\msqi{m_{\tilde{q}_i}}
\def\msqj{m_{\tilde{q}_j}}
\def\msqk{m_{\tilde{q}_k}}
\def\msql{m_{\tilde{q}_l}}
\def\mgl{m_{\tilde{g}}}

\def\as{\alpha_s}
\def\nbar{\bar{N}}
\def\bbar{\bar{b}}

\def\ca{\tilde{\chi}^\pm_1}
\def\cb{\tilde{\chi}^\pm_2}
\def\na{\tilde{\chi}^0_1}
\def\nb{\tilde{\chi}^0_2}
\def\nc{\tilde{\chi}^0_3}
\def\nd{\tilde{\chi}^0_4}
\def\ncd{\tilde{\chi}^0_{3,4}}
\def\sq{\tilde{q}}

\def\cA{{\cal A}} \def\cB{{\cal B}} \def\cC{{\cal C}} \def\cD{{\cal D}}
\def\cE{{\cal E}} \def\cF{{\cal F}} \def\cG{{\cal G}} \def\cH{{\cal H}}
\def\cI{{\cal I}} \def\cJ{{\cal J}} \def\cK{{\cal K}} \def\cL{{\cal L}}
\def\cM{{\cal M}} \def\cN{{\cal N}} \def\cO{{\cal O}} \def\cP{{\cal P}}
\def\cQ{{\cal Q}} \def\cR{{\cal R}} \def\cS{{\cal S}} \def\cT{{\cal T}}
\def\cU{{\cal U}} \def\cV{{\cal V}} \def\cW{{\cal W}} \def\cX{{\cal X}}
\def\cY{{\cal Y}} \def\cZ{{\cal Z}} 

\def\d{{\rm d}}
\def\eps{\epsilon}

\def\la{\langle }
\def\ra{\rangle }
\def\lp{\left. }
\def\rp{\right. }
\def\lr{\left( }
\def\rr{\right) }
\def\le{\left[ }
\def\re{\right] }
\def\lg{\left\{ }
\def\rg{\right\} }
\def\lb{\left| }
\def\rb{\right| }

\def\bsp#1\esp{\begin{split}#1\end{split}}

\def\beq{\begin{equation}}
\def\eeq{\end{equation}}
\def\bea{\begin{eqnarray}}
\def\eea{\end{eqnarray}}